\documentclass[10pt,twocolumn,letterpaper]{article}

\usepackage[font=small]{caption}
\usepackage{times}
\usepackage{epsfig}
\usepackage{graphicx}
\usepackage{amsmath}
\usepackage{amssymb}
\usepackage{bbm}
\usepackage{url}
\usepackage{siunitx}
\usepackage{multibib}
\newcites{suppl}{Supplement References}
\usepackage[symbol]{footmisc}

\newenvironment{myitem}
{ \begin{itemize}
    \setlength{\itemsep}{0pt}
    \setlength{\parskip}{0pt}
    \setlength{\parsep}{0pt}     }
{ \end{itemize}                } 

\DeclareMathOperator*{\argmin}{arg\,min}

\usepackage[breaklinks=true,bookmarks=false]{hyperref}

\begin{document}

\newcommand{\mytitle}{\vspace{-.5in}
\textbf{Motion Adaptive Deblurring with Single-Photon Cameras}
\vspace{-.2in}}
\title{\mytitle}

\author{
\begin{tabular}{c c c c}
  Trevor Seets$^1$ & Atul Ingle$^1$ & Martin Laurenzis$^2$ & Andreas Velten$^1$ \tabularnewline
  {\tt \small seets@wisc.edu} &{\tt\small ingle@wisc.edu} & {\tt \small martin.laurenzis@isl.eu}& {\tt \small velten@wisc.edu} 
\vspace{0.1in}
\end{tabular}\\
$^1$University of Wisconsin-Madison\\
$^2$French-German Research Institute of Saint-Louis
}
\date{}

\maketitle

\begin{abstract}
\textit{Single-photon avalanche diodes (SPADs) are a rapidly developing image sensing
technology with extreme low-light sensitivity and picosecond timing resolution.
These unique capabilities have enabled SPADs to be used in applications like
LiDAR, non-line-of-sight imaging and fluorescence microscopy that require
imaging in photon-starved scenarios. In this work we harness these capabilities
for dealing with motion blur in a passive imaging setting in low illumination
conditions. Our key insight is that the data captured by a SPAD array camera
can be represented as a 3D spatio-temporal tensor of photon detection events
which can be integrated along arbitrary spatio-temporal trajectories with
dynamically varying integration windows, depending on scene motion. We propose
an algorithm that estimates pixel motion from photon timestamp data and
dynamically adapts the integration windows to minimize motion blur. Our
simulation results show the applicability of this algorithm to a variety of
motion profiles including translation, rotation and local object motion. We
also demonstrate the real-world feasibility of our method on data captured
using a $32\times32$ SPAD camera.}
\end{abstract}

\section{Introduction}
When imaging dynamic scenes with a conventional camera, the finite exposure
time of the camera sensor results in motion blur.  This blur can be due to
motion in the scene or motion of the camera. One solution to this problem is to
simply lower the exposure time of the camera.  However, this leads to noisy
images, especially in low light conditions. In this paper we propose a technique
to address the fundamental trade-off between noise and motion blur due to 
scene motion during image capture. We focus on the challenging scenario of
capturing images in low light, with fast moving objects. Our method
relies on the strengths of rapidly emerging single-photon sensors such as single-photon avalanche diodes (SPADs).

\begin{figure}[!t]
\centering
\includegraphics[width=0.99\columnwidth]{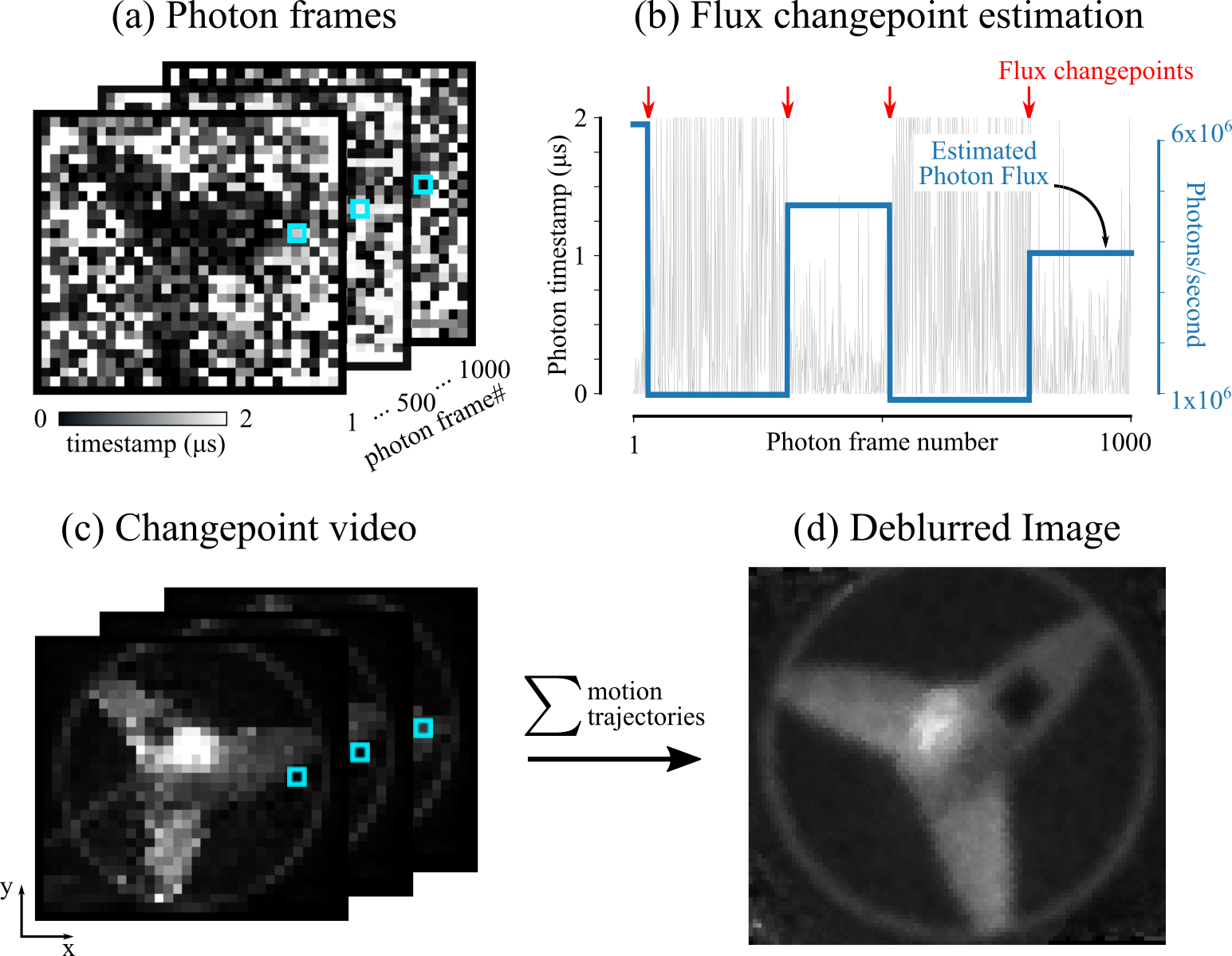}
\caption{\textbf{Overview of our adaptive motion deblurring method.}
  (a) The photon frames captured from a rotating fan contain timestamps for the
  first detected photon in each frame. Note the smaller value of timestamps in brighter regions
  and vice versa. (b) Flux changepoints obtained after
  changepoint detection and flux estimation at pixel location highlighted in cyan in (a).
  Note the rapidly varying photon timestamps due to heavy-tailed nature of the
  raw timestamp data. (c) A changepoint video (CPV) maintains sharp edges while
  providing motion cues to estimate inter-frame motion trajectories; the
  temporal flux profile of the highlighted pixel in this CPV is the piecewise
  constant function shown in (b). (d) Integrating photons along estimated
  spatio-temporal motion trajectories generates a sharp image
  with high signal-to-noise ratio.
  \label{fig:fluxchangepoints}}
\end{figure}

Light is fundamentally discrete and can be measured in terms of photons.
Conventional camera pixels measure brightness by first converting the incident
photon energy into an analog quantity (e.g.  photocurrent, or charge) that is
then measured and digitized. When imaging in low light levels, much of the
information present in the incident photons is lost due to electronic noise
inherent in the analog-to-digital conversion and readout process. Unlike
conventional image sensor pixels that require 100's-1000's of photons to
produce a meaningful signal, SPADs are sensitive down to individual photons. A
SPAD pixel captures these photons at an instant in time, with a time resolution
of hundreds of picoseconds. Each photon detection can therefore be seen as an
instantaneous event, free from any motion blur. Recently, single photon sensors have been shown to be useful when imaging in low light, or equivalently, imaging at high
frame rates where each image frame is photon-starved \cite{QBP_sizhuo20}.

The data captured by a SPAD camera is thus quite different than a conventional
camera: in addition to the two spatial dimensions, we also capture data on a
high resolution time axis resulting in a 3D spatio-temporal tensor of photon
detection events. We exploit this novel data format to deal with the noise-blur
trade-off.  Our key observation is that the photon timestamps can be combined
dynamically, across space and time, when estimating scene brightness.  If the
scene motion is known \emph{a priori}, photons can be accumulated along
corresponding spatio-temporal trajectories to create an image with no motion
blur. A conventional camera does not have this property because at capture time
high frequency components are lost \cite{raskar2006coded}. So even with known scene motion, image
deblurring is an ill-posed problem. 
 
Our method relies on dynamically changing exposure times, where each pixel can
have its own set of exposure times. An overview of our approach is shown in
Fig.~\ref{fig:fluxchangepoints}. We propose using statistical changepoint
detection to infer points in time where the photon flux at a given pixel
changes from one steady state rate to another. This allows us to
choose exposure times that adapt to scene motion. Changepoints allow us to track
high contrast edges in the scene and align ``high frequency'' motion details
across multiple image frames. We show that for the case of global motion (e.g.,
rotation) we can track the varying motion speeds of different scene pixels and
combine photons spatially to create deblurred images. We also show that for the
case of local scene motion, our method can provide robust estimates of flux
changepoints around the edges of the moving objects that improves deblurring
results obtained from downstream motion alignment and integration.

The locations of changepoints can be thought of as a spike-train generated by a
neuromorphic event camera \cite{gallego2019event}, but with three key
differences: First, unlike an event camera, our method preserves original
intensity information for each pixel. Second, the numbers and locations of
events, chosen by our algorithm, adapt to each pixel, without the need for a
hard-coded change threshold. Third, direct photon measurements are inherently
more sensitive, less noisy, and have a higher time resolution than conventional
cameras. Current cameras create an analog approximation of the quantized
photon stream (usually a charge level in a capacitor) and then measure and
digitize that analog quantity. This process introduces noise and does not take
advantage of the inherent quantized properties of the photons. We therefore
expect photon counting methods to have higher sensitivity, accuracy, and
temporal resolution than analog cameras, when imaging in low light scenarios.

\medskip
\noindent\textbf{Single-Photon Sensitive Cameras for Machine Vision?}
Currently, single-photon sensitive image sensors are quite limited in their
spatial resolution compared to conventional complementary
metal–oxide–semiconductor (CMOS) and charge-coupled device (CCD) image sensors.
However, single-photon image sensing is a rapidly developing field; recent work
has demonstrated the feasibility of making megapixel resolution single-photon
sensitive camera arrays
\cite{morimoto2020megapixel,gnanasambandam2019megapixel}.  Moreover, silicon
SPAD arrays are amenable to manufacturing at scale using the same
photolithographic fabrication techniques as conventional CMOS image sensors.
This means that many of the foundries producing our cellphone camera sensors
today could also make SPAD array sensors at similar cost. Other current
limitations of SPAD sensors are the small fraction of chip area sensitive to
light (fill factor) resulting from the large amount of additional circuitry
that is required in each pixel. Emerging 3D stacking technologies could
alleviate this problem by placing the circuitry behind the pixel.

\smallskip
\noindent{\bf Limitations}
Our experimental demonstration uses a first generation commercial SPAD array
that is limited to a 32$\times$32 pixel spatial resolution. More recently
256$\times$256 pixel commercial arrays have become available
\cite{dhulla2019256}. Although current SPAD cameras cannot compete with the
image quality of commercial CMOS and CCD cameras, with the rapid development of
SPAD array technology, we envision our techniques could be applied to future
arrays with spatial resolution similar to that of existing CMOS cameras.

\medskip
\noindent\textbf{Contributions:}
\begin{myitem}
\item We introduce the notion of \emph{flux changepoints}; these can be estimated
  using an off-the-shelf statistical changepoint detection algorithm.
\item We show that flux changepoints enable inter-frame motion estimation while
  preserving edge details when imaging in low light and at high speed.
\item We show experimental demonstration using data acquired from a
  commercially available SPAD camera.
\end{myitem}

\section{Related Work}
\noindent \textbf{Motion Deblurring}
Motion deblurring for existing cameras can be performed using blind
deconvolution~\cite{levin2008motion}. Adding a fast shutter (``flutter
shutter'') sequence can aid this deconvolution task~\cite{raskar2006coded}. We
push the idea of a fluttered shutter to the extreme limit of individual
photons: our image frames consist of individual photon timestamps allowing
dynamic adaptation of sub-exposure times for the shutter function.
Our deblurring method is inspired by burst photography pipelines used for
conventional CMOS cameras. Burst photography relies on combining frames
captured with short exposure times~\cite{hasinoff2016burst}, resulting in large amounts of data that suffer from added readout noise. Moreover,
conventional motion deblurring methods give optimal performance when the exposure time is 
matched to the true motion speed which is not known a priori.

\smallskip
\noindent\textbf{Event-based Vision Sensors}
Event cameras directly capture temporal changes in intensity instead of
capturing scene brightness \cite{gallego2019event}. Although it is possible to
create intensity images from event data in post-processing
\cite{bardow2016simultaneous}, our method natively captures scene intensities
at single-photon resolution: the ``events'' in our sensing modality are
individual photons. The notion of using photon detections as ``spiking events''
has also been explored in the context of biologically inspired vision sensors
\cite{zhu2020retina,Afshar2020eventIntensity}.  We derive flux changepoints
from the high-resolution photon timestamp data. Due to the single-photon
sensitivity, our method enjoys lower noise in low light conditions, and
pixel-level adaptivity for flux changepoint estimation.

\smallskip
\noindent\textbf{Deblurring Methods for Quanta Image Sensors}
There are two main single-photon detection technologies for passive imaging:
SPADs \cite{laurenzis2019single,ingle2019high} and quanta image sensors (QIS)
\cite{fossum2016quanta}. Although our proof-of-concept uses a SPAD camera, our
idea of adaptively varying exposure times can be applied to QIS data as well.
Existing motion deblurring algorithms for QIS
\cite{Gyongy17,gyongy2018single,QBP_sizhuo20} rely on a fixed integration
window to sum the binary photon frames. However, the initial step of picking
the size of this window requires \textit{a priori} knowledge about the motion
speed and scene brightness. Our technique is therefore complementary to existing
motion deblurring algorithms. For example, our method can be considered as a
generalization of the method in \cite{Gyngy2015BitplanePT} which uses two
different window sizes. Although we use a classical correlation-based method
for motion alignment, the sequence of flux changepoints generated using our
method can be used with state-of-the-art align-and-merge algorithms
\cite{QBP_sizhuo20} instead.


%

\section{SPAD Image Formation Model}
SPADs are most commonly used in synchronization with an active light source
such as a pulsed laser for applications including LiDAR and fluorescence
microscopy.  In contrast, here we operate the SPAD passively and only collect
ambient photons from the scene. In this section, we describe the imaging model
for a single SPAD pixel collecting light from a fixed scene point whose
brightness may vary as a function of time due to camera or scene motion.

\subsection{Pixelwise Photon Flux Estimator}
Our imaging model assumes a frame-based readout mechanism: each SPAD pixel in a
\emph{photon frame} stores at most one timestamp of the first captured ambient
photon. This is depicted in Fig.~\ref{fig:fluxchangepoints}(a).  Photon frames
are read out synchronously from the entire SPAD array; the data can be read out
quite rapidly allowing photon frame rates of 100s of kHz (frame times on the
order of a few microseconds).

Let $N_\text{pf}$ denote the number of photon frames, and $T_\text{pf}$ be the
frame period. So the total exposure time is given by $T =
N_\text{pf}T_\text{pf}$.  We now focus on a specific pixel in the SPAD array.
In the $i^\text{th}$ photon frame ($1\leq i \leq N_\text{pf}$), the output of
this pixel is tagged with a photon arrival timestamp $t_i$ relative to the
start of that frame.\footnote{In practice, due to random timing jitter and finite
resolution of timing electronics, this timestamp is stored as a discrete
fixed-point value. The SPAD camera used in our experiments has a 250 picosecond
discretization.} If no photons are detected during a photon frame, we
assume $t_i=T_\text{pf}$.  

Photon arrivals at a SPAD pixel can be modeled as a Poisson process
\cite{Hasinoff2014}. It is possible to estimate the intensity of this process
(i.e. the perceived brightness at the pixel) from the sequence of photon
arrival times \cite{laurenzis2019single,ingle2019high}. The maximum likelihood
brightness estimator $\widehat\Phi$ for the true photon flux $\Phi$ is given by
\cite{laurenzis2019single}:
\begin{equation}
    \widehat{\Phi} = \frac{\sum_{i=1}^{N_\text{pf}}\mathbf{1}(t_i \neq T_\text{pf})}{q \sum_{i=1}^{N_\text{pf}} t_i}  \label{eq:MLEFlux}
\end{equation}
where $q$ is the sensor's photon detection efficiency and $\mathbf{1}$ denotes
a binary indicator variable.

This equation assumes that the pixel intensity does not change for the exposure
time $T$. The assumption is violated in case of scene motion. If we can
determine the temporal locations of intensity changes, we can use still use
Eq.~(\ref{eq:MLEFlux}) to estimate a time-varying intensity profile for each
pixel.  In the next section we introduce the idea of \emph{flux changepoints}
and methods to locate them with photon timestamps.

\subsection{Flux Changepoints \label{ChangePoints}}
Photon flux at a given pixel may change over time in case of scene motion.
This makes it challenging to choose pixel exposure times \emph{a priori}:
ideally, for pixels with rapidly varying brightness, we should use a shorter
exposure time and vice versa. We propose an algorithm that dynamically adapts
to brightness variations and chooses time-varying exposure times on a per
pixel basis. Our method relies on locating temporal change locations where the
pixel's photon flux has a large change: we call these \emph{flux
changepoints}.

In general, each pixel in the array can have different numbers and locations of
flux changepoints. For pixels that maintain constant
brightness over the entire capture period (e.g. pixels in a static background),
there will be no flux changepoints detected and we can integrate photons over
the entire capture time $T$. For pixels with motion, we assume that the intensity between flux changepoints is constant, and we call these regions \emph{virtual exposures}. Photons in each virtual
exposure are aggregated to create a piecewise constant flux estimate. The length of each virtual
exposure will depend on how quickly
the local pixel brightness varies over time, which in turn, depends on the
true motion speed.

An example is shown in Fig.~\ref{fig:fluxchangepoints}(b). Note that the photon
timestamps are rapidly varying and extremely noisy due to a heavy-tailed
exponential distribution. Our changepoint detection algorithm detects flux
changepoints (red arrows) and estimates a piecewise constant flux waveform for
the pixel (blue plot). In the example shown, five different flux levels are
detected.

\subsubsection{Changepoint Detection}
Detecting changepoints is a well studied problem in the statistics literature
\cite{basseville1993detection, tartakovsky2014sequential}.  The goal is to
split a time-series of measurements into regions with similar statistical
properties. Here, our time series is a sequence of photon timestamps at a pixel
location, and we would like to find regions where the timestamps have the same
mean arrival rate, i.e., the photon flux during this time is roughly constant.

Using the sequence of photon arrival times $\{t_i\}_{i=1}^{N_\text{pf}}$, we
wish to find a subset $\{t_{l_1}, \ldots ,t_{l_L}\}$ representing the flux
changepoints. For convenience, we let the first and the last flux changepoint be
the first and the last photons captured by the pixel ($l_1:=1$ and $l_L =
N_\text{pf}$).

Offline changepoint detection is non-causal, i.e., it uses the full sequence of
photon timestamps for a pixel to estimate flux changepoints at that
pixel. In Suppl. Note \ref{suppl:changepoint} we describe an online
changepoint detection method that can be used for real-time applications.

For a sequence of photon timestamps, we solve the following optimization problem
\cite{rupturesPaper} (see Suppl. Note \ref{suppl:changepoint}):
\begin{equation}
    (l_i^*)_{i=1}^L\! =\! \argmin_{l_1,\!\ldots,\! l_L}
     \sum^{L-1}_{i=1}\! \left[ -\log(\widehat{\Phi}_{i}) \sum^{l_{i+1}}_{j=l_i}\mathbf{1}(t_j\! \neq\! T_\text{pf}) \right]
        \! +\! \lambda L \label{eq:ChangepointEq}
\end{equation}
where $\widehat{\Phi}_{i}$ is the photon flux estimate given by
Eq.~(\ref{eq:MLEFlux}) using only the subset of photons between times $t_{l_i}$
and $t_{l_{i+1}}$. Here $\lambda$ is the penalty term that prevents overfitting
by penalizing the number of flux changepoints. For larger test images we use
the \textsc{BottomUp} algorithm \cite{rupturesPython} which is approximate but
has a faster run time. For lower resolution images we use the Pruned
Exact Linear Time (\textsc{PELT}) algorithm \cite{PELT} which gives an exact
solution but runs slower. See Suppl. Section \ref{suppl:bottomup_v_pelt} for results using the \textsc{BottomUp} algorithm.

\smallskip
\noindent{\bf Flux Changepoints for QIS Data}
The cost function in Eq.~(\ref{eq:ChangepointEq}) applies to photon timestamp
data, but the same idea of adaptive flux changepoint detection can be used with
QIS photon count data as well.  A modified cost function that uses photon
counts (0 or 1) is derived in Suppl. Note~\ref{suppl:qis_changepoint}.

\begin{figure}[!t]
    \centering
    \includegraphics[width=\columnwidth]{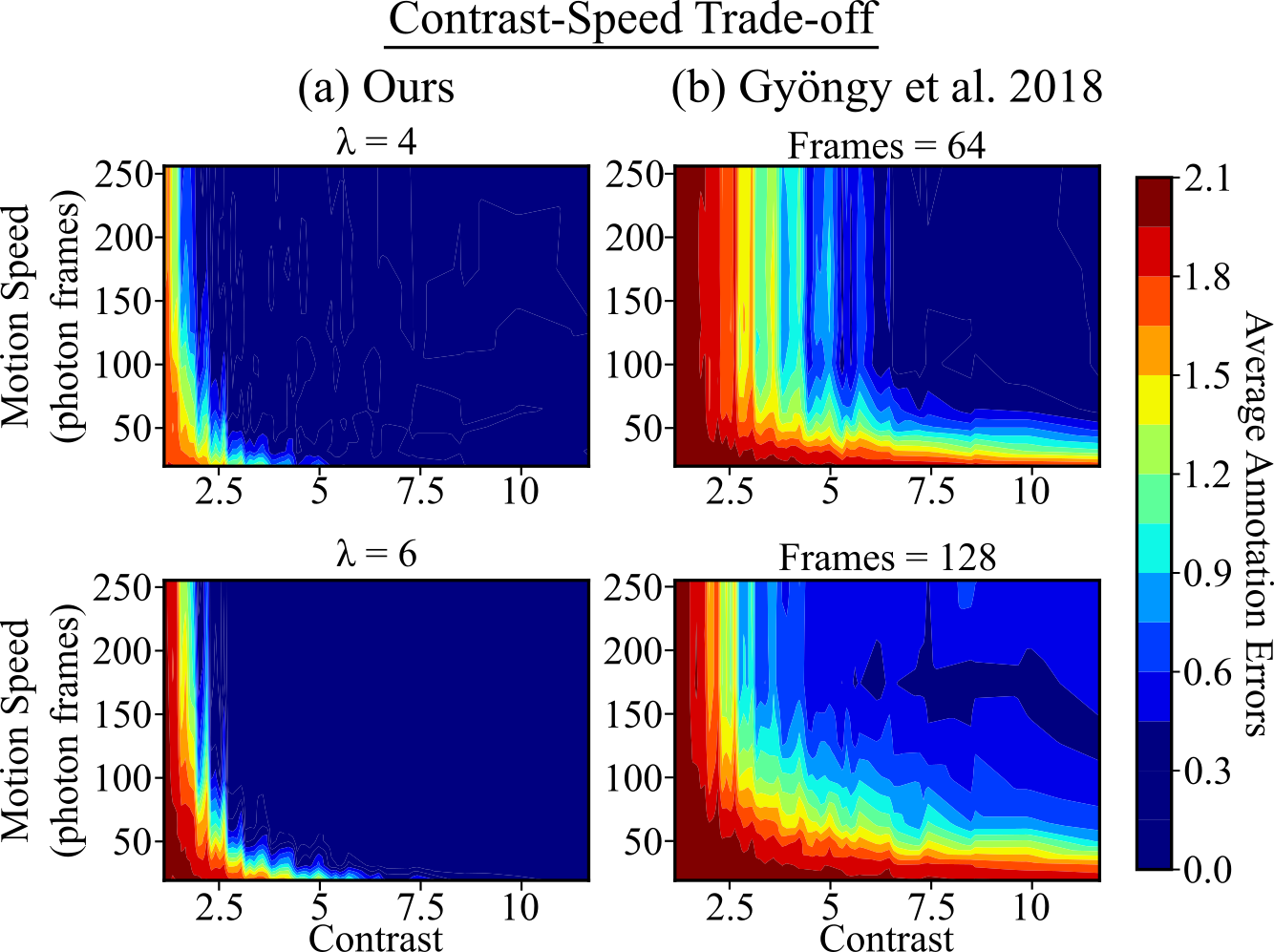}
    \caption{{\bf Contrast-Speed Trade-off.}  We simulate a single SPAD pixel
    measuring a pulse-shaped photon flux signal to analyze the contrast-speed
    trade-off when detection motion. Our method (a) based on PELT changepoint detection
    adapts to a wider range of contrast and speed combinations than the
    comparison method (b) from Gy\"ongy et al. \cite{gyongy2018single}.
    \label{fig:contrast_vs_speed}}
\end{figure}

\subsubsection{Single-Pixel Simulations}
There is a trade-off between contrast and motion speed when trying to detect
changepoints. For example, it is harder to detect the flux change with a fast
moving object with a lower contrast with respect to its background.  To
evaluate this trade-off, we simulate a single SPAD pixel for 800 photon frames
with a time varying flux signal with a randomly placed pulse wave. We use the
pulse width as a proxy for motion speed, and vary the contrast by varying the
ratio of the pulse height.  We measure the absolute difference between the
number of changepoints detected by our algorithm and the true number of flux
changes (which is exactly 2 for a single pulse).  We call this ``annotation
error.''

Fig.~\ref{fig:contrast_vs_speed} shows the annotation errors for different
values of contrast and motion speeds for two different algorithms. For each set
of contrasts and motion speeds we display the average number of annotation errors over 120 simulation runs. We use the \textsc{PELT} algorithm to
detect flux changepoints. For comparison, we also show the fixed
windowing approach used by Gy{\"o}ngy et al.
\cite{gyongy2018single,Agresti_Coull_Method}.  The changepoint detection
algorithm is able to adapt to a wider range of contrasts and speeds.



\begin{figure*}[!ht]
\centering
\includegraphics[width=\linewidth]{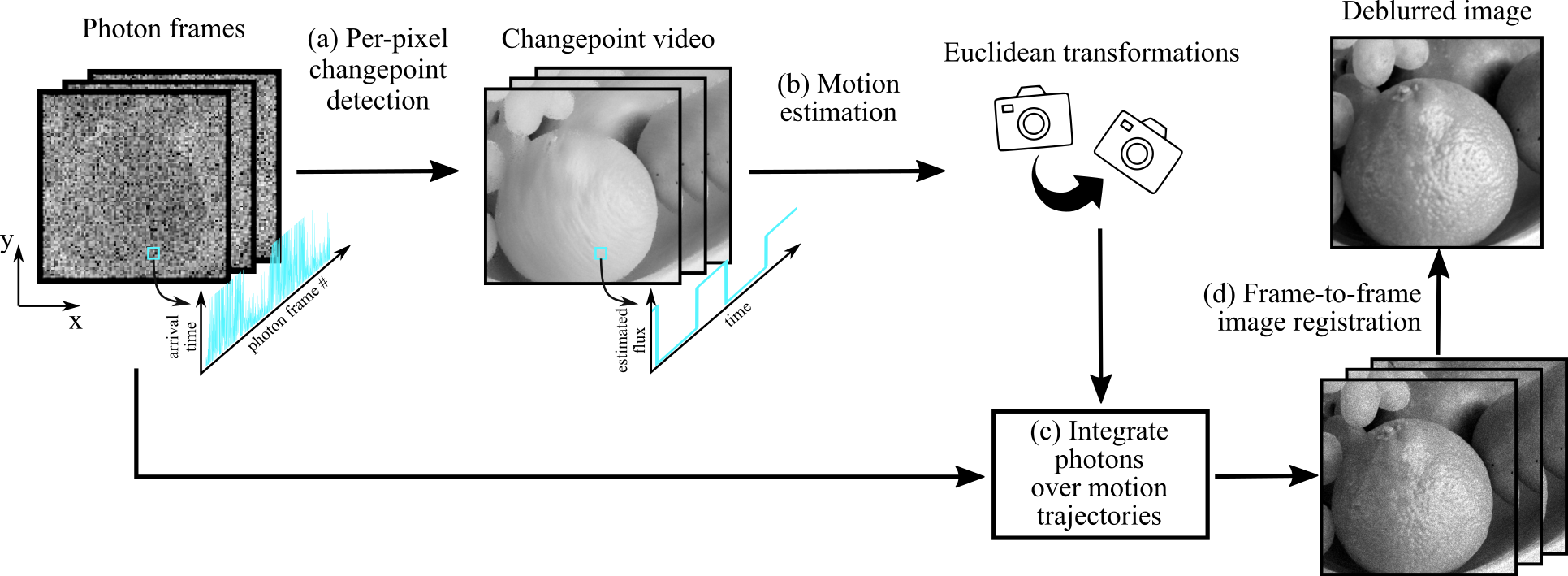}
\caption{\textbf{Changepoint video deblurring}. Photon timestamps for each
  pixel location in the 3D timestamp tensor are analyzed for changepoints (a) to
  generate a changepoint video. Pixels may have different numbers of
  changepoints, and hence, dynamically changing frame-rate over time. We estimate
  motion between successive frames of the changepoint video (b) and estimate
  motion parameters. Finally, using this motion estimate, we sum photons (c) from
  the photon frames along motion trajectories resulting in a deblurred video.
  Frames in the deblurred video are registered and summed (d) to obtain the final
  deblurred image. (Original images from FreeImages.com) \label{fig:flowchart} }
\end{figure*}

\section{Pixel-Adaptive Deblurring\label{sec:deblurring}}
\subsection{Changepoint Video}
Using methods described in the previous section we locate flux changepoints for
each pixel in the image. The changepoints for each pixel represent when a new
virtual exposure starts and stops; photons within each virtual exposure can be
used to estimate photon flux using Eq.~(\ref{eq:MLEFlux}).  We call this
collection of piecewise constant functions over all pixels in the array the
\emph{changepoint video} (CPV).

The CPV does not have an inherent frame rate; since each pixel has a continuous
time piecewise constant function, it can be sampled at arbitrarily spaced time
instants in $[0,T]$ to obtain any desired frame rate. We sample the CPV at
non-uniform time intervals using the following criterion. Starting with the
initial frame sampled at $t=0$, we sample the subsequent frames at instants
when at least 1\% of the pixel values have switched to a new photon flux. This
leads to a variable frame rate CPV that adapts to changes in scene velocity:
scenes with fast moving objects will have a higher average CPV frame rate.


The CPV preserves large brightness changes in a scene. For example, the edges
of a bright object moving across a dark background remain sharp. However, finer
texture details within an object may appear blurred.



\subsection{Spatio-Temporal Motion Integration}
We use global motion cues from the CPV and then combine the motion estimates
with the photon frames to create a deblurred video. The overall flowchart is
shown in Fig.~\ref{fig:flowchart}.

\begin{figure}[!t]
  \centering \includegraphics[width=\columnwidth]{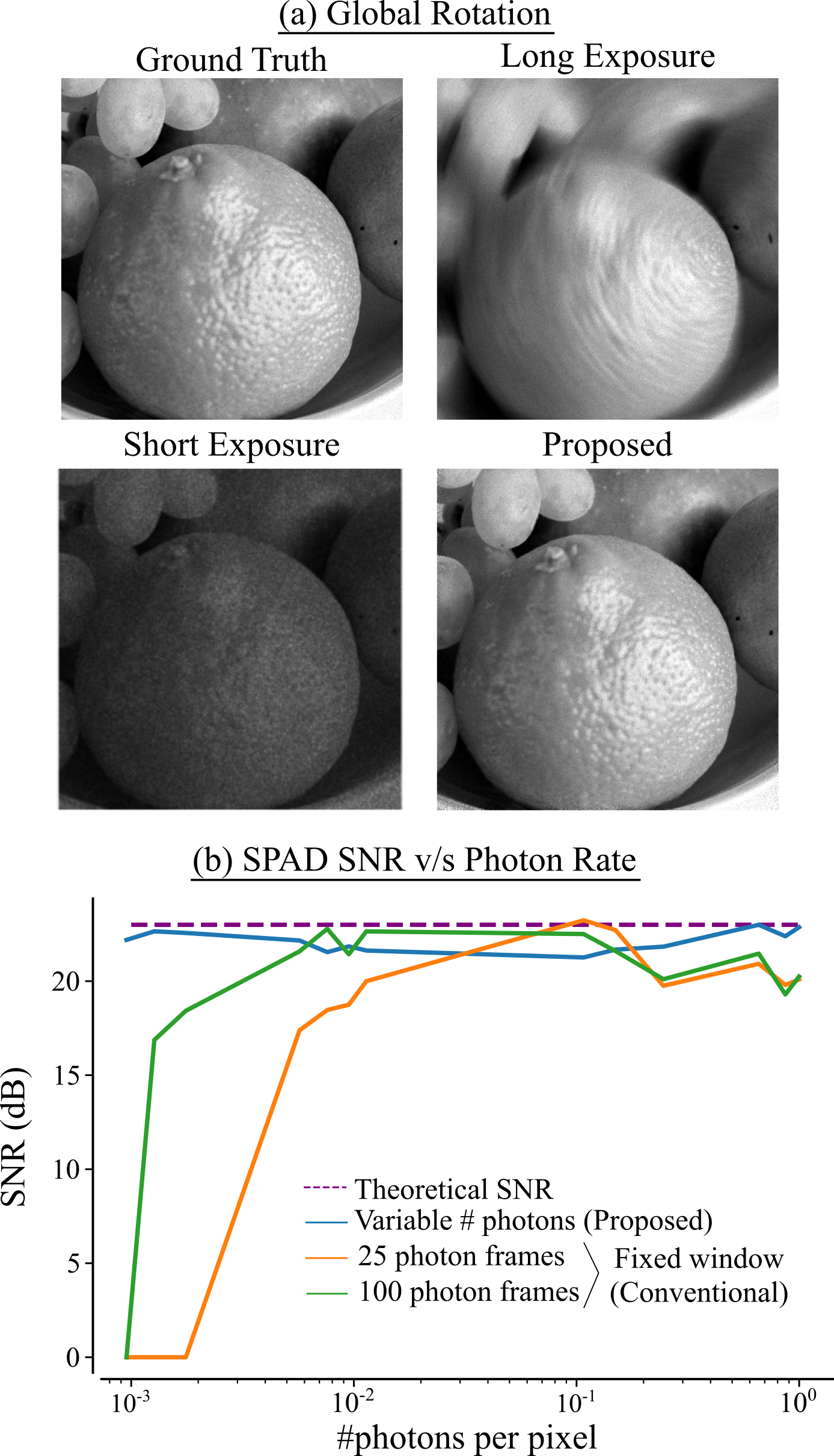}
  \caption{
    \textbf{Comparison of SNR for different deblurring window sizes.}
    (a) A ground truth video of a rotating orange used for creating
    simulated SPAD photon frames. A long exposure image is quite blurry while a short
    exposure image is very noisy. Our deblurring algorithm strikes a balance between noise
    and blur to get a sharp high-quality image. (b) By varying the brightness
    of the orange scene, we compare the simulated SNR using our method compared
    to a conventional method with fixed windows.
    Notice the proposed method stays above 20 dB at all photon rates,
    while the fixed photon window SNRs decrease in the low photon count regime.
    (Original image from FreeImages.com)
    \label{fig:PhotonBurstSize}}
\end{figure}

We use a correlation-based image registration algorithm to find the motion
between consecutive frames in the CPV. For each location $\mathbf{p}=[x,y]$ in
the $i^\text{th}$ CPV frame, we assume it maps to a location  $\mathbf{p}'$ in
$(i+1)^\text{st}$ CPV frame. We assume that the mapping is a linear
transformation:
\begin{equation}
    A\mathbf{p}=\mathbf{p}'.\label{registrationEq}
\end{equation}
We constrain $A$ to represent planar Euclidean motion (rotation and
translation). Let $\theta$ be the amount of rotation centered around a point
$[r_x,r_y]$ and let $[\tau_x,\tau_y]$ be the translation vector. Then $A$ has
the form:
\begin{equation}
    A\!=\!\begin{bmatrix}
\cos\theta & \sin\theta & r_x(1\!-\!\cos\theta)\!+\!r_y\sin\theta \!+\!\tau_x\\
-\sin\theta & \cos\theta & r_y(1\!-\!\cos\theta)\!-\!r_x\sin\theta\!+\!\tau_y\\
0&0&1
\end{bmatrix}.
\end{equation}
We apply the enhanced correlation coefficient maximization algorithm
\cite{evangelidis2008parametric} to estimate the transformation matrix
$A$ for consecutive pairs of CPV frames $i \rightarrow i+1$.
A sequence of frame-to-frame linear transformations generates arbitrarily
shaped global motion trajectories. We aggregate the original photon frame data along these estimated spatio-temporal motion trajectories.

We assume that the rotation center, $[r_x,r_y]$, is the middle of the
image and a change in rotation center can be modeled as a translation. We solve for $\theta$, $\tau_x$, and $\tau_y$ which we linearly
interpolate. Then using the interpolated motion parameters and Eq.~(\ref{registrationEq}), we align all photon
frames corresponding to the time interval between CPV frames $i\rightarrow i+1$ and sum these frames to get a photon flux
image by using Eq.~(\ref{eq:MLEFlux}) at each pixel. This generates a motion
deblurred video with the same frame rate as the CPV, but with finer textures
preserved as shown in Fig.~\ref{fig:flowchart}.

If the final goal is to obtain a single deblurred image, we repeat the steps
described above on consecutive frames in the deblurred video, each time
decreasing the frame rate by a factor of 2, until eventually we get a single
image. This allows us to progressively combine photons along spatial-temporal
motion trajectories to increase the overall signal to noise ratio (SNR) and
also preserve high frequency details that were lost in the CPV.

Our method fails when motion becomes too large to properly align images,
especially at low resolutions. It can also fail when not many flux changepoints
are detected, this will occur mainly due to a lack of photons per pixel of
movement. In the worst case, if not enough changepoints are detected, the
result of the algorithm will look similar to a single long exposure image. 

The method of aligning and adding photon frames is similar to contrast
maximization algorithms used for event cameras \cite{gallego2018unifying}.
However, unlike event camera data, our method relies on the CPV
which contains both intensity and flux changepoints derived from
single-photon timestamps.

\medskip
\noindent{\bf Handling Multiple Moving Objects}
To handle multiple moving objects on a static background, we implement a method
similar to Gy{\"o}gny et al. \cite{gyongy2018single} and combine it with our CPV method.  We cluster the changepoints for different objects using a
density-based spatial clustering algorithm (DBSCAN) \cite{esterDBSCAN}. For
each cluster, we then create a bounding box, isolating
different moving objects. We then apply our motion deblurring algorithm on each
object individually, before stitching together each object with the areas that
are not moving in the CPV.  The clustering step also denoises by rejecting flux changepoints not belonging to a cluster.
\subsection{Simulations}\label{sec:simulation}
Starting with a ground truth high resolution, high frame rate video, we scale
the video frames to units of photons per second and generate photon frames
using exponentially distributed arrival times. We model a photon frame readout SPAD array with \num{8000} bins and bin width of \SI{256}{\ps}. 

We first simulate a rotating orange by applying successive known rigid
transformations to an image and generating SPAD data by scaling the transformed
images between $10^4$ and $10^8$ photons per second. We rotate the image by 0.1$^\circ$
for every 10 generated photons for a total of 1000 photons. We use the
\textsc{BottomUp} algorithm \cite{rupturesPaper} with $\lambda=5$ for the
changepoint detection step. The results are shown in
Fig.~\ref{fig:PhotonBurstSize}(a). Our method captures sharp details on the orange
skin while maintaining high quality.

Fig.~\ref{fig:PhotonBurstSize}(b) shows quantitative comparisons of SNR for different
deblurring methods. The conventional approach to deblur photon data uses a fixed frame
rate; we use two different window lengths for comparison.  We compute the SNR
of the deblurred imaging using the $\ell 2$ (root-mean-squared) distance from the
ground truth to and repeat this over a range of photon flux levels.  We keep
the total number of photons captured approximately constant by extending the
capture time for darker flux levels. Our method dynamically adapts to motion
and lighting so we are able to reconstruct with high SNR even in photon starved
regimes where the SNR of the fixed window methods degrades rapidly.

\begin{figure*}[!ht]
  \centering \includegraphics[width=.95\linewidth]{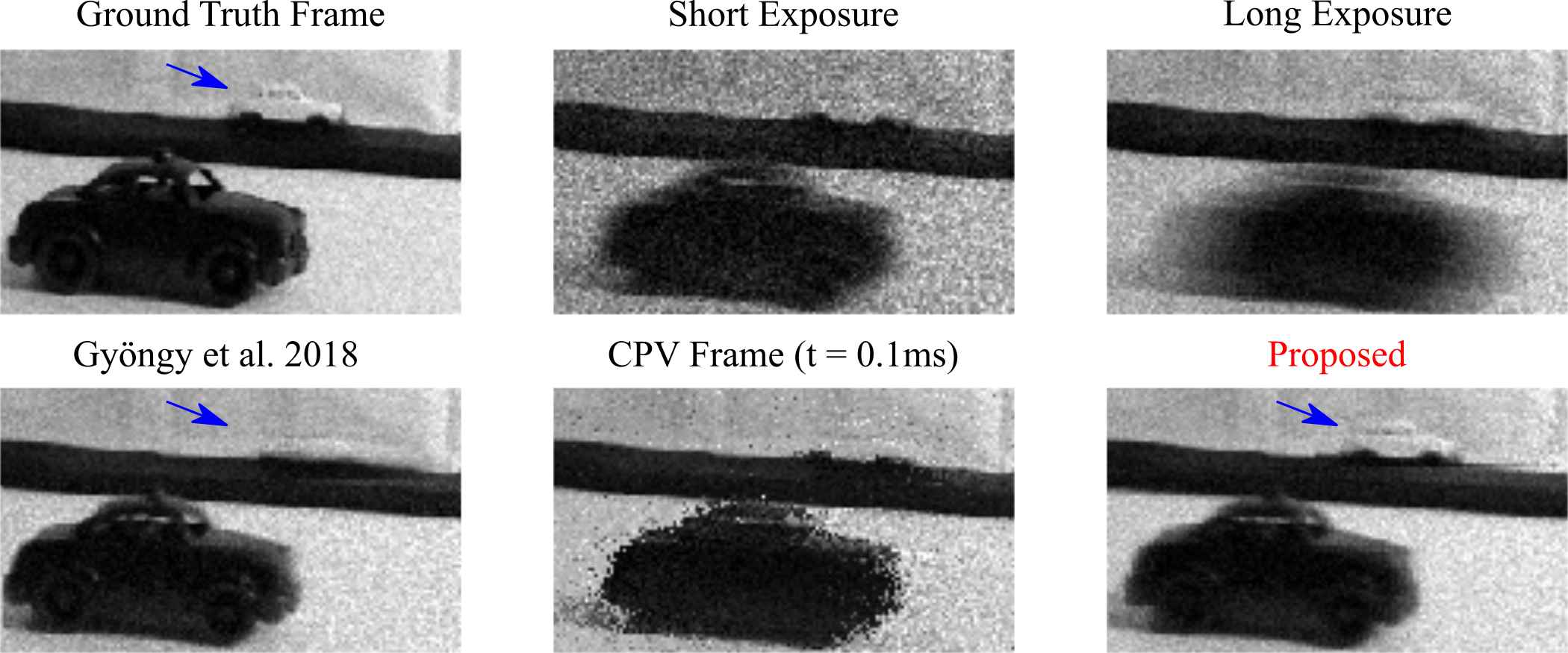}
  \caption{
    \textbf{Simulated Motion Deblurring for Multiple Moving Objects.} We simulate SPAD
    data from video of toy cars, a fast moving black car and slow moving white
    car. (Top row) A sample frame from the ground truth frame sequence is
    shown.  The short and long exposure images show the results of using
    integrating the first 75 and 250 photon frames, respectively.  Notice that
    the short exposure preserves the black car while the white car is quite
    noisy, on the other hand, the long average blurs the black car but preserves
    details of the white one better. (Bottom row) The method of Gy{\"o}gny et
    al. \cite{gyongy2018single}, fails to reconstruct the white car (blue arrow) due to its
    low contrast.  A sample frame from our changepoint video shows both moving
    cars. Finally, our deblurring algorithm is able to reconstruct both the
    black and white car with negligible motion blur.
    \label{fig:sim_cars}}
\end{figure*}

\begin{figure}[!ht]
 \centering \includegraphics[width=\columnwidth]{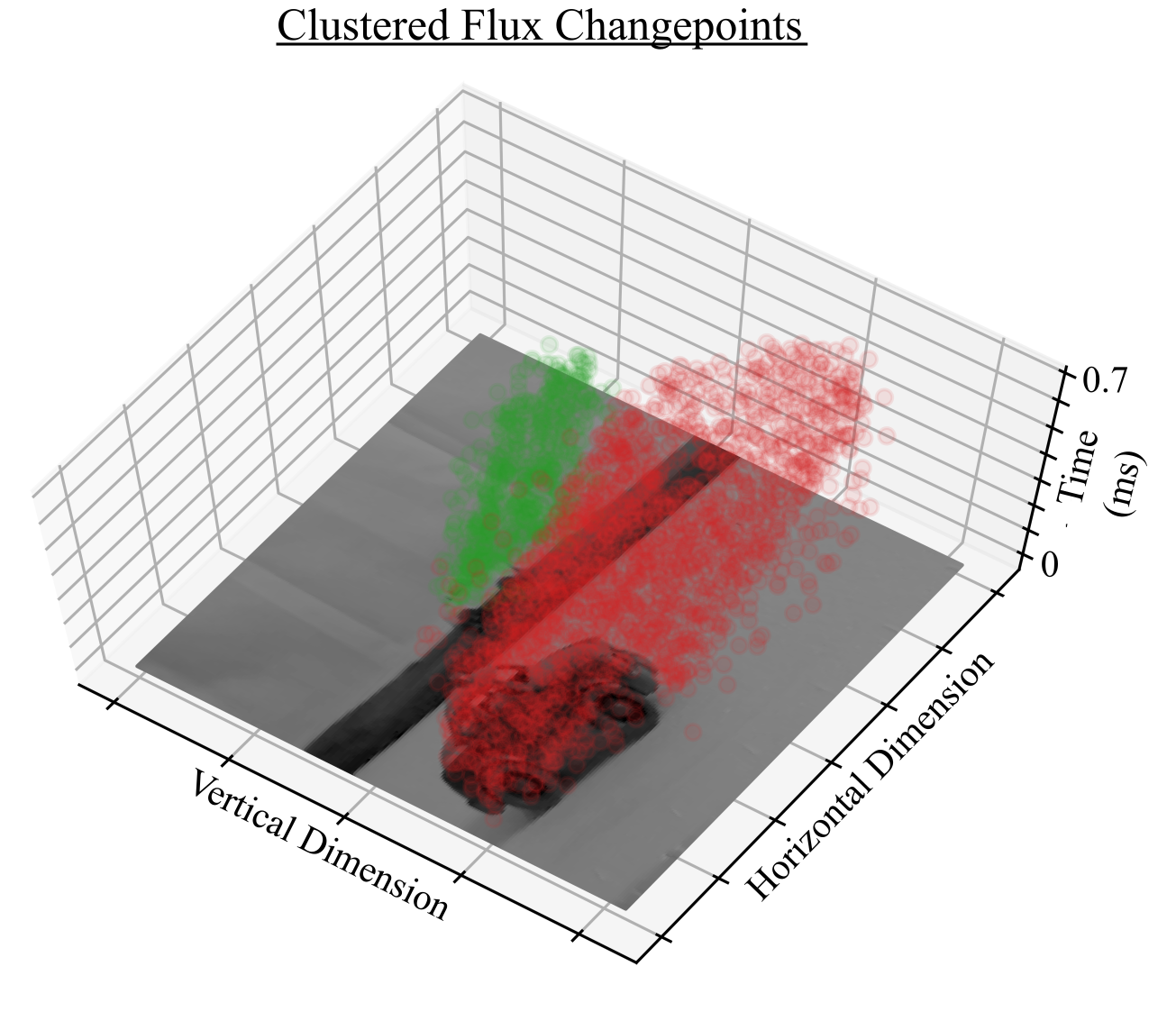}
  \caption{
    \textbf{Clustered Flux Changepoints} Two clusters of flux changepoints
    are detected using frames from the changepoint video for the toy car scene.
    These changepoint clusters are used for segmenting the moving cars from the
    static background.
    \label{fig:clustered_changepoints}}
    \vspace{-0.20in}
\end{figure}

To simulate multi-object motion, we captured a ground truth video of two toy
cars rolling down a ramp at different speeds. The video frame pixels are then
scaled between $10^5$ and $10^6$ photons per second and a total of 690 photon
frames are generated. A bright slow moving car has a contrast of 1.2 with
respect to the background, and moves 48 pixels over the duration of video. The
dark car has a contrast of 5.0 with the background, and moves 143 pixels. We
use the \textsc{PELT} algorithm \cite{PELT} with $\lambda=6$ for the
changepoint detection step. We use $\epsilon = 7.5$ and $\text{MinPts} = 40$ in
the DBSCAN clustering algorithm.  The resulting deblurred images are shown in
Fig.~\ref{fig:sim_cars}. Observe that the method of \cite{gyongy2018single}
blurs out the low contrast white car in the back. Our method assigns
dynamically changing integration windows extracted from the CPV to successfully
recover both cars simultaneously with negligible motion blur. The changepoint clusters used for segmenting cars from the static background in our method are are shown in Fig.~\ref{fig:clustered_changepoints}. 


\begin{figure*}[!ht]
    \centering
    \includegraphics[width=\linewidth]{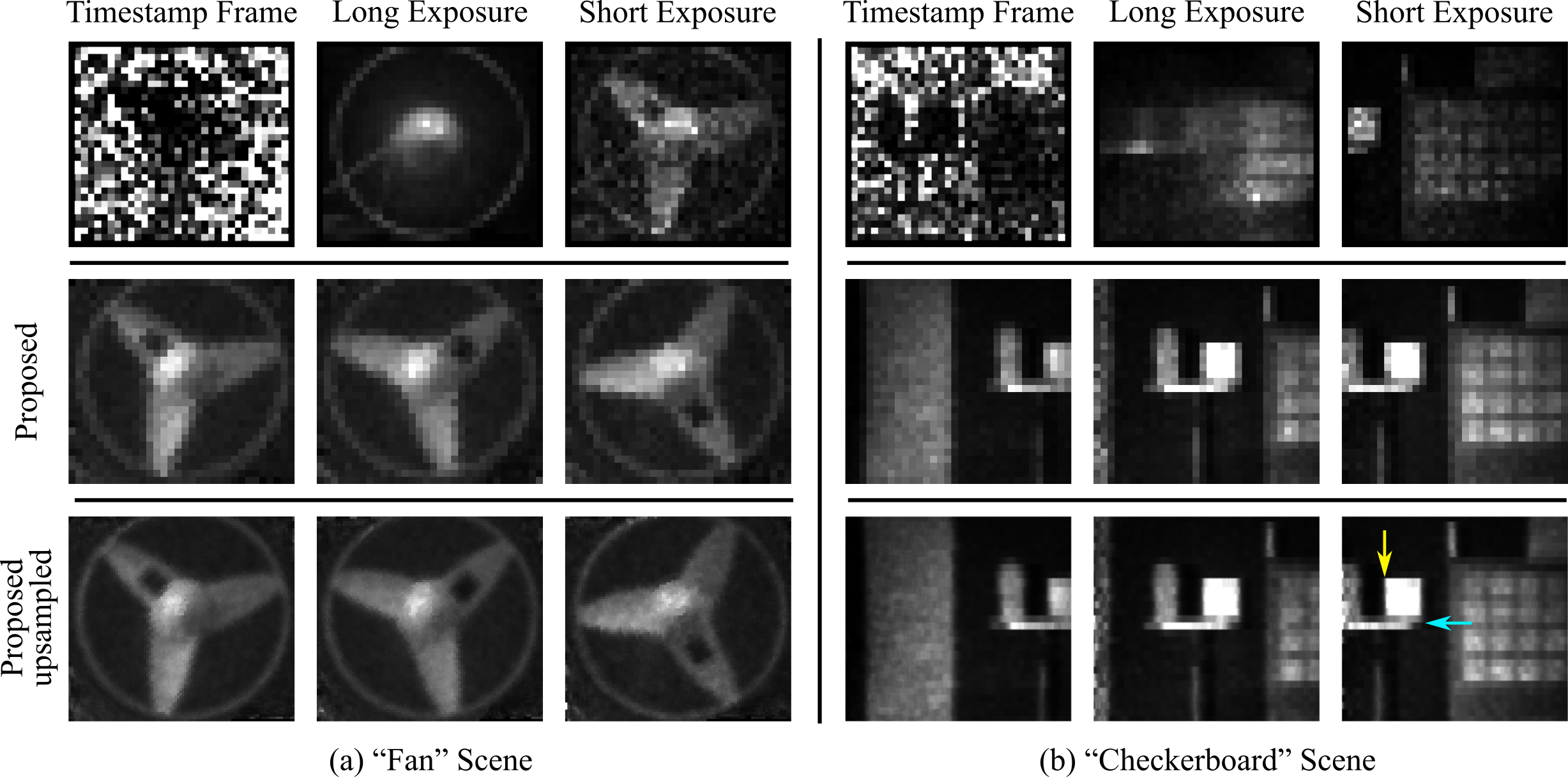}
    \caption{{\bf Experimental Results.} The first row shows a single raw data
    frame from the photon timestamp tensor; each photon has an associated timestamp with a
    \SI{250}{\pico\second} bin resolution. Integration over a long exposure
    (\SI{2}{\ms} for the fan and \SI{0.2}{\ms} for
    checkerboard scene) this gives a low noise but blurry result. Using a short
    exposure time (\SI{40}{\micro\second} for both scenes) produces very noisy results.
    The second row shows a sequence of three frames from the final deblurred video.
    The third row shows the result with upsampling. Note that in the checkerboard scene due to purely
    horizontal motion, some vertical edges (yellow arrow) are sharper but not
    the horizontal edges (cyan arrow).  See supplementary video results.
    \label{fig:expt_result} \vspace{-0.10in}
  }
\end{figure*}

\section{Experiments}
We validate our method using experimental data captured using a 32$\times$32
InGaAs SPAD array from Princeton Lightwave Inc., USA. The pixels are sensitive
to near infrared and shortwave infrared (\SI{900}{\nm}--\SI{1.6}{\micro\meter}
wavelengths). The SPAD array operates in a frame readout mode at 50,000 photon
frames per second. Each photon frame exposure window is \SI{2}{\micro\second}
and sub-divided into 8000 bins, giving a temporal resolution of
\SI{250}{\pico\second} per bin \cite{PLCamDatasheet}.

For this low resolution experimental data, we zero-order hold upsample both the
timestamp frames and the CPV before step (b) in Fig. \ref{fig:flowchart} in the
spatial dimensions. We then use upsampled photon frames for step (c) resulting
in sharper images. Upsampling before motion integration allows photons that are
captured in the same pixel to land in a larger space during the motion
integration step.

Fig.~\ref{fig:expt_result} shows results from two different scenes. The ``fan''
scene shows the performance of our algorithm with fast rotation\footnote[2]{The
background behind the fan is covered with absorbing black felt material. This
allows us to treat the data as having global rotation, because there is hardly
any light captured from the background.}. The optimal exposure time in this
case depends on the rotation speed. Our method preserves the details of the fan
blades including the small black square patch on one of the fan blades.
The ``checkerboard'' scene shows deblurring result with purely horizontal
global motion. Note that our method is able to resolve details such as the
outlines of the squares on the checkerboard.

The last row in Fig.~\ref{fig:expt_result} shows the upsampled results.
Benefits of upsampling are restricted to the direction of motion. The ``fan''
dataset is upsampled 9$\times$ compared to the original resolution. The
``checkerboard'' dataset is upsampled 4$\times$; this is
because the motion is limited to the horizontal dimension. Note
that some of the details in the vertical edges are sharp, but horizontal edges
remain blurry.



\section{Discussion and Future Work}
\noindent\textbf{Euclidean Motion Assumption}
In the case of camera shake, most of the motion will come from small rotations
in the camera that result in 2D translations and rotations in the image frames.
The short distance translations of shaking camera would cause translations in
the frames that are similar in nature, but smaller in magnitude.

Translation of the camera over larger distances would result in parallax while
motion within the scene can result in more complex changes. In these cases our
model captures scene changes only approximately. It applies to frame to frame
motion over short time-scales and limited to regions in the scene. Ideas for
extending our model to deal with larger motion will be the subject of future work.

\smallskip
\noindent\textbf{Dealing with Local Motion} 
The techniques presented in this paper assume multiple moving objects
exhibiting euclidean motion, with no occlusions. We can extend our approach to
more complex motions. We can use a patch-wise alignment and merging methods to
deal with more complex local motion and occlusions \cite{hasinoff2016burst,
QBP_sizhuo20}. 

Deblurring algorithms developed for event camera data can be adapted to SPAD
data, because the flux changepoints represent changes in brightness similar to
the output of event cameras.  Current event camera algorithms are able to
recover complex motion in scenes
\cite{gallego2018unifying,stoffregenEventSegmentation}, and they could be
improved with a fusion based approach where image intensity information is also
available \cite{Gehrig19ijcv}.

\smallskip
\noindent\textbf{Data Compression}
With increasing number of pixels, processing photon frames with high
spatio-temporal resolution will be quite resource intensive. Our online
changepoint method takes some initial steps towards a potential real-time
implementation. The CPV can be used for video compression with variable frame rate: by tuning the regularization parameter of the changepoint detection
algorithm a tradeoff between image fidelity and data rate can be achieved.

\footnotesize{\smallskip
\noindent\textbf{Acknowledgments}
This material is based upon work supported by the Department of Energy / National Nuclear Security Administration under Award Number DE-NA0003921, the National Science Foundation (GRFP DGE-1747503, CAREER 1846884), DARPA (REVEAL HR0011-16-C-0025) and the Wisconsin Alumni Research Foundation. 

\smallskip
\noindent\textbf{Disclaimer}
This report was prepared as an account of work sponsored by an agency of the United States Government.  Neither the United States Government nor any agency thereof, nor any of their employees, makes any warranty, express or implied, or assumes any legal liability or responsibility for the accuracy, completeness, or usefulness of any information, apparatus, product, or process disclosed, or represents that its use would not infringe privately owned rights.  Reference herein to any specific commercial product, process, or service by trade name, trademark, manufacturer, or otherwise does not necessarily constitute or imply its endorsement, recommendation, or favoring by the United States Government or any agency thereof.  The views and opinions of authors expressed herein do not necessarily state or reflect those of the United States Government or any agency thereof.}

{\small
\bibliographystyle{ieee_fullname}
\bibliography{egbib}

\begin{thebibliography}{10}\itemsep=-1pt

\bibitem{Afshar2020eventIntensity}
S. {Afshar}, T.~J. {Hamilton}, L. {Davis}, A. {Van Schaik}, and D. {Delic}.
\newblock Event-based processing of single photon avalanche diode sensors.
\newblock {\em IEEE Sensors Journal}, pages 1--1, 2020.

\bibitem{Agresti_Coull_Method}
Alan Agresti and Brent~A. Coull.
\newblock Approximate is better than "exact" for interval estimation of
  binomial proportions.
\newblock {\em The American Statistician}, 52(2):119--126, 1998.

\bibitem{bardow2016simultaneous}
Patrick Bardow, Andrew~J Davison, and Stefan Leutenegger.
\newblock Simultaneous optical flow and intensity estimation from an event
  camera.
\newblock In {\em Proceedings of the IEEE Conference on Computer Vision and
  Pattern Recognition}, pages 884--892, 2016.

\bibitem{basseville1993detection}
Mich{\`e}le Basseville, Igor~V Nikiforov, et~al.
\newblock {\em Detection of abrupt changes: theory and application}, volume
  104.
\newblock Prentice Hall Englewood Cliffs, 1993.

\bibitem{dhulla2019256}
Vinit Dhulla, Sapna~S Mukherjee, Adam~O Lee, Nanditha Dissanayake, Booshik Ryu,
  and Charles Myers.
\newblock 256 x 256 dual-mode cmos spad image sensor.
\newblock In {\em Advanced Photon Counting Techniques XIII}, volume 10978, page
  109780Q. International Society for Optics and Photonics, 2019.

\bibitem{esterDBSCAN}
Martin Ester, Hans-Peter Kriegel, J\"{o}rg Sander, and Xiaowei Xu.
\newblock A density-based algorithm for discovering clusters in large spatial
  databases with noise.
\newblock In {\em Proceedings of the Second International Conference on
  Knowledge Discovery and Data Mining}, KDD'96, page 226–231. AAAI Press,
  1996.

\bibitem{evangelidis2008parametric}
Georgios~D Evangelidis and Emmanouil~Z Psarakis.
\newblock Parametric image alignment using enhanced correlation coefficient
  maximization.
\newblock {\em IEEE Transactions on Pattern Analysis and Machine Intelligence},
  30(10):1858--1865, 2008.

\bibitem{fossum2016quanta}
Eric~R Fossum, Jiaju Ma, Saleh Masoodian, Leo Anzagira, and Rachel Zizza.
\newblock The quanta image sensor: Every photon counts.
\newblock {\em Sensors}, 16(8):1260, 2016.

\bibitem{gallego2019event}
G. {Gallego}, T. {Delbruck}, G.~M. {Orchard}, C. {Bartolozzi}, B. {Taba}, A.
  {Censi}, S. {Leutenegger}, A. {Davison}, J. {Conradt}, K. {Daniilidis}, and
  D. {Scaramuzza}.
\newblock Event-based vision: A survey.
\newblock {\em IEEE Transactions on Pattern Analysis and Machine Intelligence},
  pages 1--1, 2020.

\bibitem{gallego2018unifying}
Guillermo Gallego, Henri Rebecq, and Davide Scaramuzza.
\newblock A unifying contrast maximization framework for event cameras, with
  applications to motion, depth, and optical flow estimation.
\newblock {\em Proceedings / CVPR, IEEE Computer Society Conference on Computer
  Vision and Pattern Recognition. IEEE Computer Society Conference on Computer
  Vision and Pattern Recognition}, 06 2018.

\bibitem{Gehrig19ijcv}
Daniel Gehrig, Henri Rebecq, Guillermo Gallego, and Davide Scaramuzza.
\newblock {EKLT}: Asynchronous, photometric feature tracking using events and
  frames.
\newblock {\em Int. J. Comput. Vis.}, 2019.

\bibitem{gnanasambandam2019megapixel}
Abhiram Gnanasambandam, Omar Elgendy, Jiaju Ma, and Stanley~H Chan.
\newblock Megapixel photon-counting color imaging using quanta image sensor.
\newblock {\em Optics express}, 27(12):17298--17310, 2019.

\bibitem{Gyongy17}
I. Gy{\"o}ngy, T.~A. Abbas, N. Dutton, and R. Henderson.
\newblock Object tracking and reconstruction with a quanta image sensor.
\newblock In {\em Proceedings of the International Image Sensor Workshop},
  2017.

\bibitem{gyongy2018single}
Istvan Gyongy, Neale~AW Dutton, and Robert~K Henderson.
\newblock Single-photon tracking for high-speed vision.
\newblock {\em Sensors}, 18(2):323, 2018.

\bibitem{Hasinoff2014}
Samuel~W. Hasinoff.
\newblock {\em Photon, Poisson Noise}, pages 608--610.
\newblock Springer US, Boston, MA, 2014.

\bibitem{hasinoff2016burst}
Samuel~W Hasinoff, Dillon Sharlet, Ryan Geiss, Andrew Adams, Jonathan~T Barron,
  Florian Kainz, Jiawen Chen, and Marc Levoy.
\newblock Burst photography for high dynamic range and low-light imaging on
  mobile cameras.
\newblock {\em ACM Transactions on Graphics (TOG)}, 35(6):1--12, 2016.

\bibitem{ingle2019high}
Atul Ingle, Andreas Velten, and Mohit Gupta.
\newblock High flux passive imaging with single-photon sensors.
\newblock In {\em Proceedings of the IEEE Conference on Computer Vision and
  Pattern Recognition}, pages 6760--6769, 2019.

\bibitem{Gyngy2015BitplanePT}
Gyongy Istvan, Dutton Neale, Luca Parmesan, Davies Amy, Saleeb Rebecca, Duncan
  Rory, Rickman Colin, Dalgarno Paul, and Robert~K Henderson.
\newblock Bit-plane processing techniques for low-light, high speed imaging
  with a spad-based qis.
\newblock In {\em International Image Sensor Workshop}, pages 1--4, 2015.

\bibitem{PELT}
R. Killick, P. Fearnhead, and I.~A. Eckley.
\newblock Optimal detection of changepoints with a linear computational cost.
\newblock {\em Journal of the American Statistical Association},
  107(500):1590--1598, 2012.

\bibitem{laurenzis2019single}
Martin Laurenzis.
\newblock Single photon range, intensity and photon flux imaging with kilohertz
  frame rate and high dynamic range.
\newblock {\em Optics Express}, 27(26):38391--38403, 2019.

\bibitem{levin2008motion}
Anat Levin, Peter Sand, Taeg~Sang Cho, Fr{\'e}do Durand, and William~T Freeman.
\newblock Motion-invariant photography.
\newblock {\em ACM Transactions on Graphics (TOG)}, 27(3):1--9, 2008.

\bibitem{QBP_sizhuo20}
Sizhuo Ma, Shantanu Gupta, Arin~C. Ulku, Claudio Bruschini, Edoardo Charbon,
  and Mohit Gupta.
\newblock Quanta burst photography.
\newblock {\em ACM Trans. Graph.}, 39(4), July 2020.

\bibitem{morimoto2020megapixel}
Kazuhiro Morimoto, Andrei Ardelean, Ming-Lo Wu, Arin~Can Ulku, Ivan~Michel
  Antolovic, Claudio Bruschini, and Edoardo Charbon.
\newblock Megapixel time-gated spad image sensor for 2d and 3d imaging
  applications.
\newblock {\em Optica}, 7(4):346--354, 2020.

\bibitem{raskar2006coded}
Ramesh Raskar, Amit Agrawal, and Jack Tumblin.
\newblock Coded exposure photography: motion deblurring using fluttered
  shutter.
\newblock In {\em ACM SIGGRAPH 2006 Papers}, pages 795--804. ACM, 2006.

\bibitem{stoffregenEventSegmentation}
Timo Stoffregen, Guillermo Gallego, Tom Drummond, Lindsay Kleeman, and Davide
  Scaramuzza.
\newblock Event-based motion segmentation by motion compensation.
\newblock In {\em Proceedings of the IEEE International Conference on Computer
  Vision}, pages 7244--7253, 2019.

\bibitem{tartakovsky2014sequential}
Alexander Tartakovsky, Igor Nikiforov, and Michele Basseville.
\newblock {\em Sequential analysis: Hypothesis testing and changepoint
  detection}.
\newblock CRC Press, 2014.

\bibitem{PLCamDatasheet}
Princeton~Lightwave/AMS Technologies.
\newblock {\em 32 x 32 Geiger-mode Avalanche Photodiode (GmAPD) Camera}, 2012
  (accessed June 20, 2020).
\newblock
  \url{http://www.amstechnologies.com/fileadmin/amsmedia/downloads/4796_gmapdcameradatasheet.pdf}.

\bibitem{rupturesPython}
C. Truong.
\newblock {\em ruptures Python Package}, 2017 (accessed June 20, 2020).
\newblock
  \url{https://ctruong.perso.math.cnrs.fr/ruptures-docs/build/html/index.html}.

\bibitem{rupturesPaper}
Charles Truong, Laurent Oudre, and Nicolas Vayatis.
\newblock Selective review of offline change point detection methods.
\newblock {\em Signal Processing}, 167:107299, 2020.

\bibitem{zhu2020retina}
Lin Zhu, Siwei Dong, Jianing Li, Tiejun Huang, and Yonghong Tian.
\newblock Retina-like visual image reconstruction via spiking neural model.
\newblock In {\em Proceedings of the IEEE/CVF Conference on Computer Vision and
  Pattern Recognition}, pages 1438--1446, 2020.

\end{thebibliography}


\begin{thebibliography}{1}\itemsep=-1pt

\bibitem{Adams2007BayesianOC}
Ryan~P. Adams and David J.~C. MacKay.
\newblock Bayesian online changepoint detection.
\newblock {\em arXiv: Machine Learning}, 2007.

\bibitem{ingle2019high}
Atul Ingle, Andreas Velten, and Mohit Gupta.
\newblock High flux passive imaging with single-photon sensors.
\newblock In {\em Proceedings of the IEEE Conference on Computer Vision and
  Pattern Recognition}, pages 6760--6769, 2019.

\bibitem{onlineDetectionCode}
Johannes Kulick, nariox, Dan Marthaler, Minesh~A. Jethva, and Sean Kruzel.
\newblock bayesian changepoint detection.
\newblock \url{https://github.com/hildensia/bayesian_changepoint_detection},
  2020.

\bibitem{rupturesPaper}
Charles Truong, Laurent Oudre, and Nicolas Vayatis.
\newblock Selective review of offline change point detection methods.
\newblock {\em Signal Processing}, 167:107299, 2020.

\end{thebibliography}
}

\clearpage
\onecolumn
\normalsize
\renewcommand{\figurename}{Supplementary Figure}
\renewcommand{\thesection}{S.\arabic{section}}
\renewcommand{\theequation}{S\arabic{equation}}
\setcounter{figure}{0}
\setcounter{section}{0}
\setcounter{equation}{0}
\setcounter{page}{1}

\begin{center}
\Large Supplementary Document for\\[0.2cm]
\Large ``\mytitle'' \\[1.5cm]
\vspace{.3in}
\normalsize Trevor Seets, \ Atul Ingle,  \ Martin Laurenzis, \ Andreas Velten\\
Correspondence to: seets@wisc.edu 

\end{center}

\section{Flux Changepoint Detection \label{suppl:changepoint}}
\subsection{Offline Algorithm: Cost Function Derivation}
Consider a set of photon time stamp measurements $\{x_i\}_{i=1}^N$. Here each
$x_i$ is a valid measurement, and in the frame-readout capture mode, described
in the main text, this is different than the $t_i$'s. If no photon is detected
in a frame we add the frame length to the next detected photon. We do this so
each $x_i$ will be i.i.d. and distributed exponentially. We again wish to find
a set of change points, $\{x_{l_1}, \ldots ,x_{l_L}\}$. In general, the
optimization problem for changepoint detection is given by Eq.~(P2) in
\citesuppl{rupturesPaper}: 

\begin{equation}
    (l_i^*)_{i=1}^L\! =\! \argmin_{l_1,\!\ldots,\! l_L}
     \sum^{L-1}_{i=1}\! c(x_{l_i...l_{i+1}}).
        \! +\! \lambda L \label{eq:suppl_changePointGeneral}
\end{equation}

The summation term represents the likelihood that each segment in between
changepoints come from the same underlying distribution, while the
regularization term is needed because the number of changepoints are not known
\emph{a priori}. For our case $c(\cdot)$ is the negative log likelihood for a
set of exponentially distributed measurements. Let $f(x)$ be the exponential
density function with rate parameter $\Phi$, and let $\widehat{\Phi}_i$ be the
maximum likelihood estimate for $\Phi$ for the set of measurements
$\{x_{l_i}\ldots x_{l_{i+1}}\}$. Note that the maximum likelihood estimator
maximizes the log likelihood. To derive $c(\cdot)$, we begin with Eq. (C1) from
\citesuppl{rupturesPaper}:

\begin{align}
    c(x_{l_i...l_{i+1}}) &=-\underset{\Phi}{\max}\sum_{j=l_i}^{l_{i+1}}\log f(x_j | \Phi) \\
    &= -\sum_{j=l_i}^{l_{i+1}}\log f(x_j | \widehat{\Phi}_i)\\
    &= -\sum_{j=l_i}^{l_{i+1}}\log \widehat{\Phi}_ie^{-\widehat{\Phi}_ix_j}\\
    &= -\sum_{j=l_i}^{l_{i+1}}\log \widehat{\Phi}_i+\sum_{j=l_i}^{l_{i+1}}\widehat{\Phi}_ix_j\\
    &= -(l_{i+1}-l_i)\log \widehat{\Phi}_i+(l_{i+1}-l_i)\label{eq:suppl_costFunction}
\end{align}
where the last line comes from the fact that
$\widehat{\Phi}_i=\frac{(l_{i+1}-l_i)}{\sum_{j=l_i}^{l_{i+1}} x_j}$.
Plugging Eq.~(\ref{eq:suppl_costFunction}) into
Eq.~(\ref{eq:suppl_changePointGeneral}), the last term sums to a
constant $N$ and can be dropped from the optimization. Then we convert to the
direct measurments $t_i$ by expanding out where no photons where found to get
Eq.~(\ref{eq:ChangepointEq}).

\subsection{QIS: Offline Cost Function\label{suppl:qis_changepoint}}
A quanta image sensor (QIS) is another sensor type capable of measuring single photons. Unlike a SPAD, the QIS senor only gives a binary output for each photon-frame corresponding to whether or not a photon was detected. Note that we can convert our experimental SPAD data to QIS data by stripping the SPAD data of the timing information. Let $n_i=0$ if the $i^{th}$ QIS photon-frame detects no photons and $n_i=1$ otherwise. Let $\tau_b$ be the temporal bin width for each photon-frame. Suppose the jot is exposed to a flux of $\Phi$, then the probability of detecting a photon during photon-frame $i$ is:
\begin{equation}
    p=P(n_i=1) = 1-e^{-q\Phi\tau_b}\label{eq:suppl:QIS_prob_1}
\end{equation}
Where $q$ is the quantum efficiency. We can model measuring multiple photon-frames with the QIS jot as a Bernoulli Trial, with probability of success given by Eq. \ref{eq:suppl:QIS_prob_1}. For a set of $N$ photon-frames the maximum likelihood estimator, $\hat{\Phi}_{QIS}$, is given by \citesuppl{ingle2019high},

\begin{align}
    \hat{\Phi}_{QIS} = \frac{-1}{q\tau_b}\ln({1-\widehat{p}})\label{eq:suppl:QIS_MLE} \\
    \widehat{p}=\frac{\sum_{i=0}^Nn_i}{N}
\end{align}

This flux estimator should also be used in SPAD sensors under very high fluxes, where their is significant probability of detecting more than one photon in a period equal to the SPAD's time quantization. Similarly, the MLE in Eq. \ref{eq:MLEFlux} can be used in low light conditions for a QIS sensor. 

 We derive the changepoint cost function in the raw data domain. Following the steps of the earlier derivation, with $f(n_i)$ being the Bernoulli distribution with parameter $p$:

\begin{align}
    c_{QIS}(x_{l_i...l_{i+1}}) &=-\underset{\Phi}{\max}\sum_{j=l_i}^{l_{i+1}}\log f(n_j | \Phi) \\
    &= -\sum_{j=l_i}^{l_{i+1}}\log f(n_j | \widehat{\Phi}_i)\\
    &= -\log(\widehat{p_i}) \sum_{j=l_i}^{l_{i+1}}n_j - \log(1-\widehat{p_i}) \sum_{j=l_i}^{l_{i+1}}1-n_j\label{eq:suppl_QIS_costFunction}
\end{align}

Where $\widehat{p_i}=\frac{\sum_{j=l_i}^{l_{i+1}} n_j}{l_{i+1}-l_i}$. 

\subsection{Online Flux Changepoint Detection Algorithm \label{suppl:online}}
Offline changepoint detection is suitable for offline applications that capture
a batch of photon frames and generate a deblurred image in post-processing. In
some applications that require fast real-time feedback (e.g. live deblurred
video) or on-chip processing with limited frame buffer memory, online
changepoint detection methods can be used.  We use a Bayesian online
changepoint detection method \citesuppl{Adams2007BayesianOC}. This algorithm
calculates the joint probability distribution of the time since the last flux
changepoint. For exponentially distributed data, it uses the posterior
predictive distribution which is a Lomax distribution (see Suppl. Note
\ref{suppl:online}).  We assume that the flux changepoints appear uniformly
randomly in the exposure window $T$.  Because detecting a flux changepoint
after only one photon is difficult we use a look-behind window that evaluates
the probability of the photon 20--40 photon frames into the past as being a
flux changepoint. Using a look-behind window greatly increases detection
accuracy and introduces only minor latency (on the order of tens of
microseconds). We also found that it is helpful to use a small spatial window
that spreads out flux changepoints in space to increase the density of
changepoints. In general, online detection will work better for slower motion
as the algorithm learns from past data. We compare online and offline detection
in Suppl. Note \ref{suppl:additional_results}.

We use a Bayesian changepoint detection algorithm shown in Algorithm 1 of
\citesuppl{Adams2007BayesianOC}. Here we derive the posterior predictive distribution
used in Step 3 of their algorithm. We use a $\textsf{Gamma}(\alpha,\beta)$ prior
for $\Phi$. Let $\mathbf{x} :=\{x_i\}_{i=1}^N.$
It can be shown that $\Phi |\mathbf{x}  \sim \textsf{Gamma}(\alpha+N,
\beta+\sum_{i=1}^N x_i)$. The predictive posterior density is given by:
\begin{align}
  f_{x_{N+1}|\mathbf{x}}(y | \mathbf{x}) &= \int_0^\infty f_{x_{N+1}|\mathbf{\Phi}}(y | \Phi)
                                                   f_{\mathbf{\Phi}|\mathbf{x}}(\Phi | {\bf x}) d\Phi \\
											 &= \int_0^\infty \Phi e^{-\Phi y} \frac{(\beta + \sum x_i)^{\alpha+N}}{\Gamma(\alpha+N)}
                                                 \Phi^{\alpha+N-1} e^{-(\beta+\sum_i x_i)\Phi} d\Phi \\
                       &= \frac{(\beta + \sum_i x_i)^{\alpha+N} (\alpha+N)}{(\beta + y + \sum_i x_i)^{\alpha+N-1}}
\end{align}
which is a Lomax density with shape parameter $\alpha+N$ and scale parameter
$\beta+\sum_i x_i$. For our data we used a $\mathsf{Lomax}(1,100)$ in Step 3
and $H(\cdot) = 40$ in Steps 4 and 5 of Algorithm 1 in
\citesuppl{Adams2007BayesianOC}.

\begin{figure}[!ht]
  \centering \includegraphics[width=0.55\columnwidth]{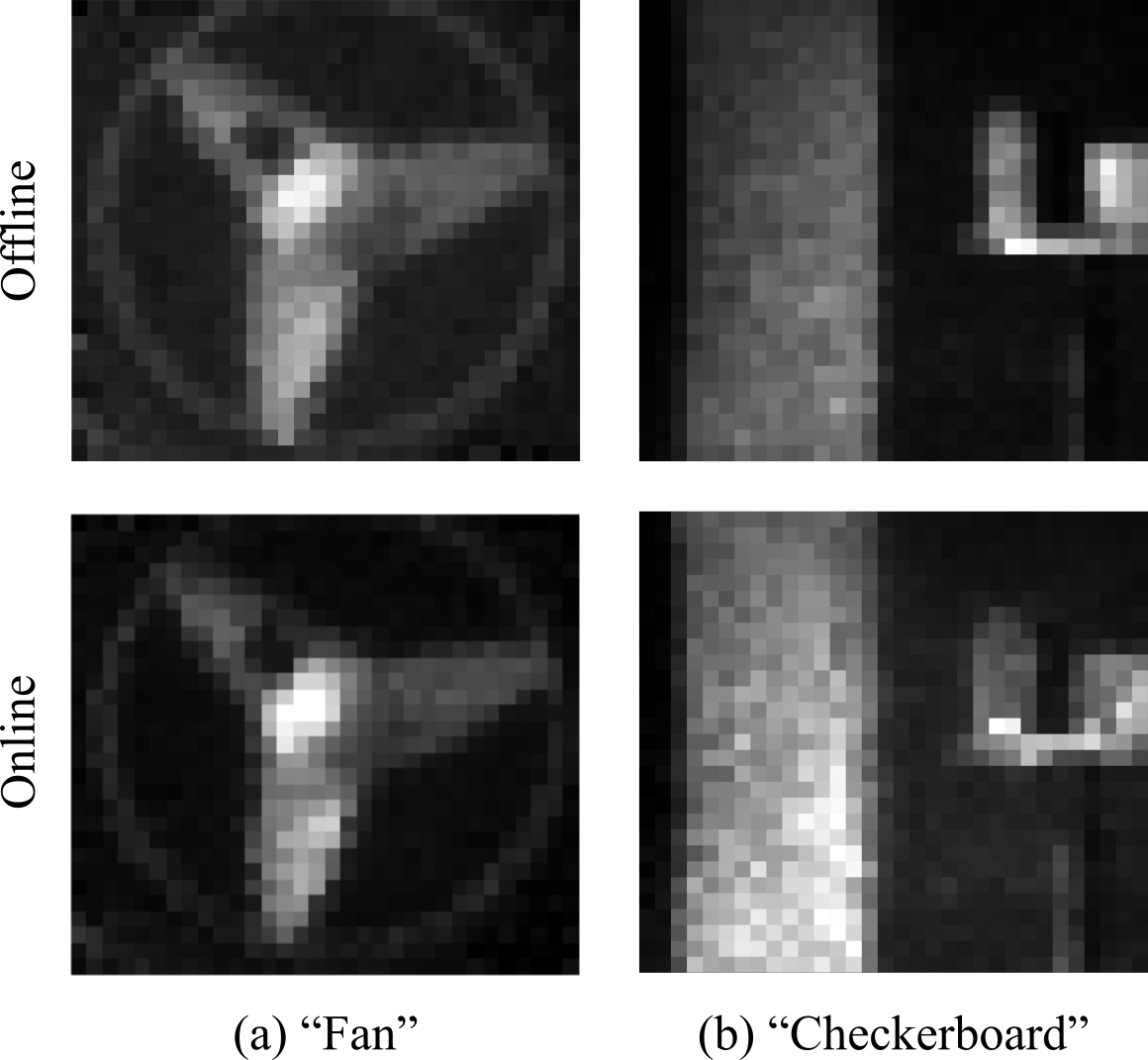}
  \caption{\textbf{Comparing online vs. offline changepoint detection.} We processed the two
  experimental datasets using our online and offline changepoint detection
  algorithms. There is a slight loss of edge details when the online algorithm
  is used. \label{fig:onlineVsOfflineExpt} }
\end{figure}

For online detection we use code modified from \citesuppl{onlineDetectionCode} (Commit: 7d21606859feb63eba6d9d19942938873915f8dc). Fig.~\ref{fig:onlineVsOfflineExpt} shows the results of using the online changepoint
detection algorithm. Observe that some of the edge details are better preserved
with the offline changepoint algorithm. 

We run the same experiment as in Fig.~\ref{fig:PhotonBurstSize}, where an orange is rotated at different brightness levels. We show the resulting SNR for online vs offline detection in supplementary Fig.~\ref{fig:suppl_online_snr}

\begin{figure*}[!ht]
  \centering \includegraphics[width=0.55\linewidth]{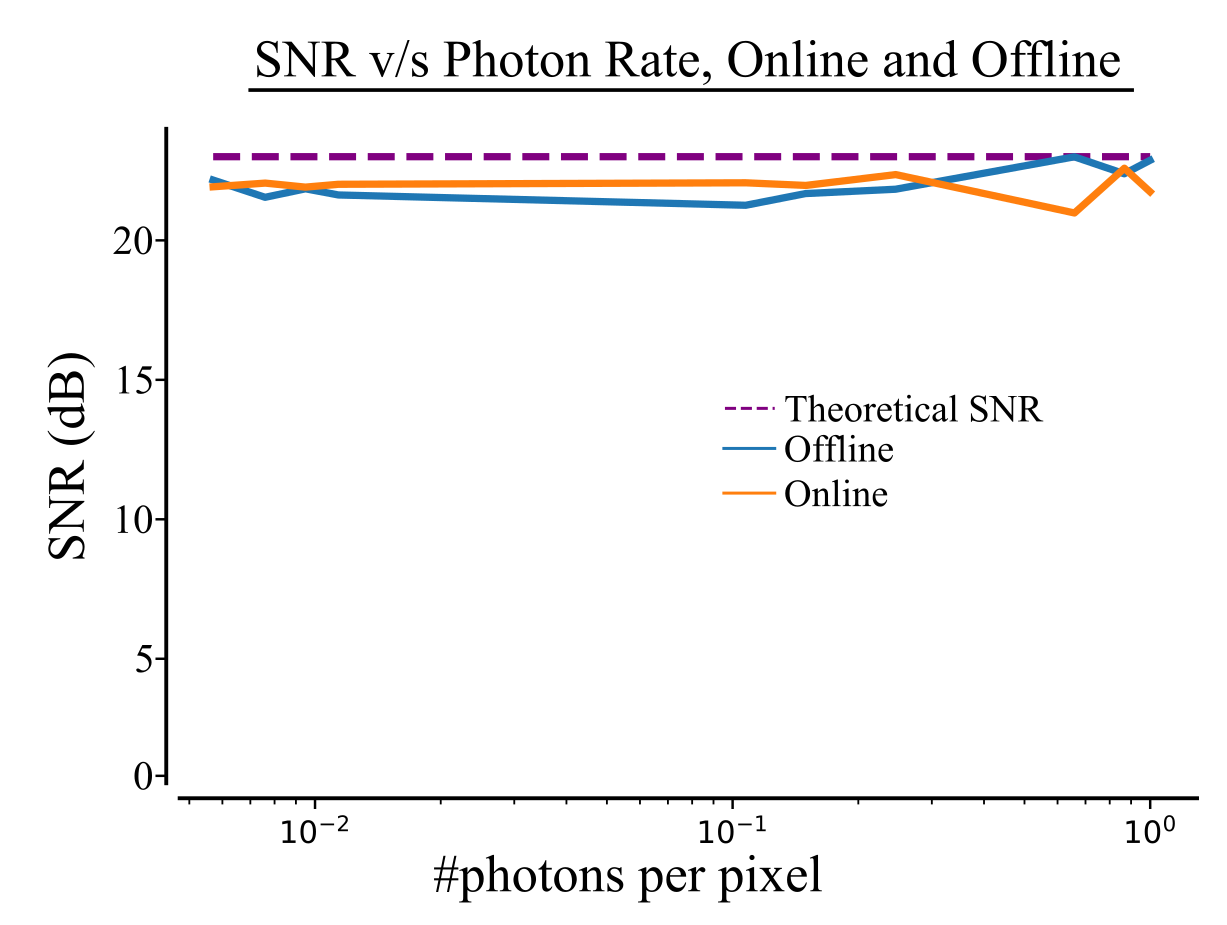}
  \caption{
   \textbf{Online vs. Offline Rotating Orange SNR} We run the same experiment as in Fig.~\ref{fig:PhotonBurstSize}, with the rotating orange at many light levels. Note that the resulting SNRs of the two methods are quite similar.
    \label{fig:suppl_online_snr}}
\end{figure*}

\clearpage
\section{SNR Analysis \label{suppl:effect_of_speed}}
\begin{figure}[!ht]
  \centering \includegraphics[width=0.5\columnwidth]{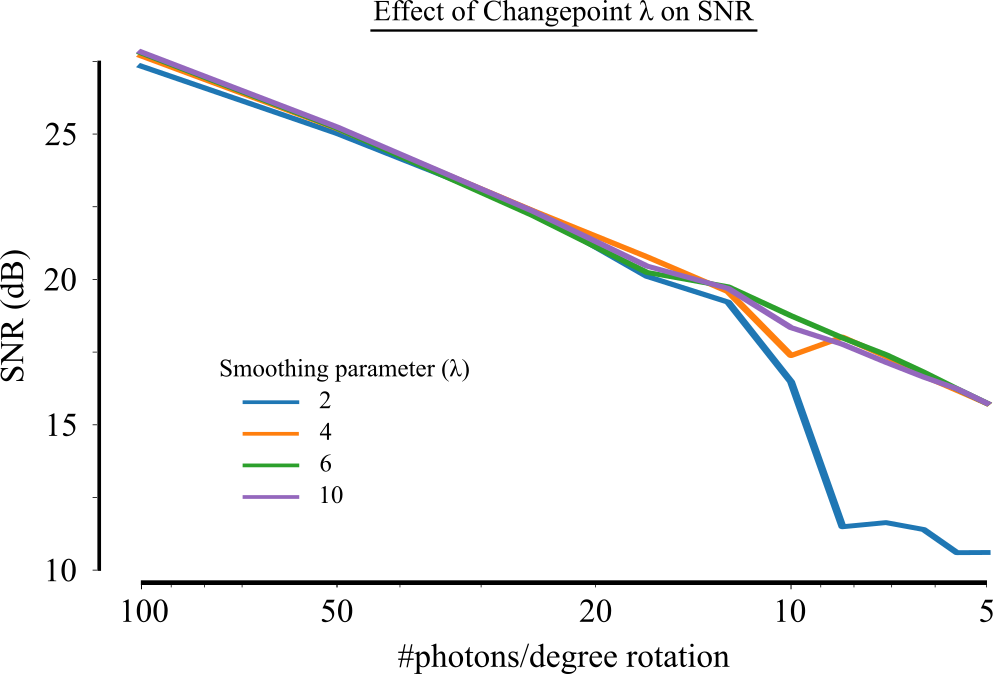}
  \caption{\textbf{Effect of offline changepoint algorithm regularization
  parameter $\lambda$.} If $\lambda$ is large enough we get good performance.
  When $\lambda$ is too small, many flux changepoints are found, which will
  cause the CPV to be too noisy to properly align frames.
  \label{fig:suppl_snrVsSmoothingParam} }
\end{figure}

\begin{figure}[!ht]
  \centering \includegraphics[width=0.5\columnwidth]{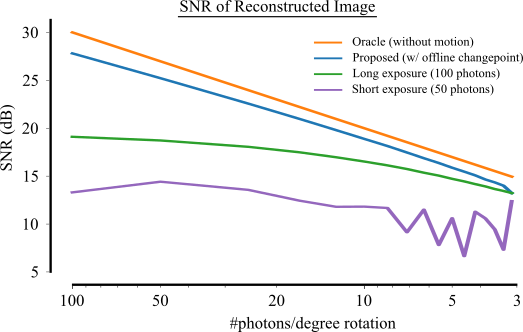}
  \caption{Effect of number of photons captured per degree of rotation. Our
    algorithm is unable to find flux changepoints at speeds of 3 photon frames
    per degree. When no flux changepoints are found we just get a long exposure
    image. We also see that the SNR for a blurry image (long exposure) or a
    noisy image (short exposure) is worse than the proposed deblurring method
    until our method converges to the long exposure image.
    \label{fig:suppl_SNRvsSpeed}}
\end{figure}

We test our deblurring algorithm for different motion speeds for the case of
rotational motion using the ``orange'' dataset in the main text. We measure the
SNR by computing the discrepancy between the ground truth flux image and the
deblurred result. We do this by temporally downsampling the original photon
frames, so the number of photons decrease as the motion speeds up.
Suppl.~Fig.~\ref{fig:suppl_snrVsSmoothingParam} shows how changing the
regularization parameter $\lambda$ in the offline flux changepoint detection
algorithm effects the SNR. We find that as long as $\lambda$ is high enough a
good reconstruction SNR stays high.  In Suppl.~Fig.~\ref{fig:suppl_SNRvsSpeed}
we show that our algorithm converges to the performances of a long exposure
capture (with motion blur) if the number of photons per degree of rotation
falls below 3.

\begin{figure}[!ht]
  \centering \includegraphics[width=0.5\linewidth]{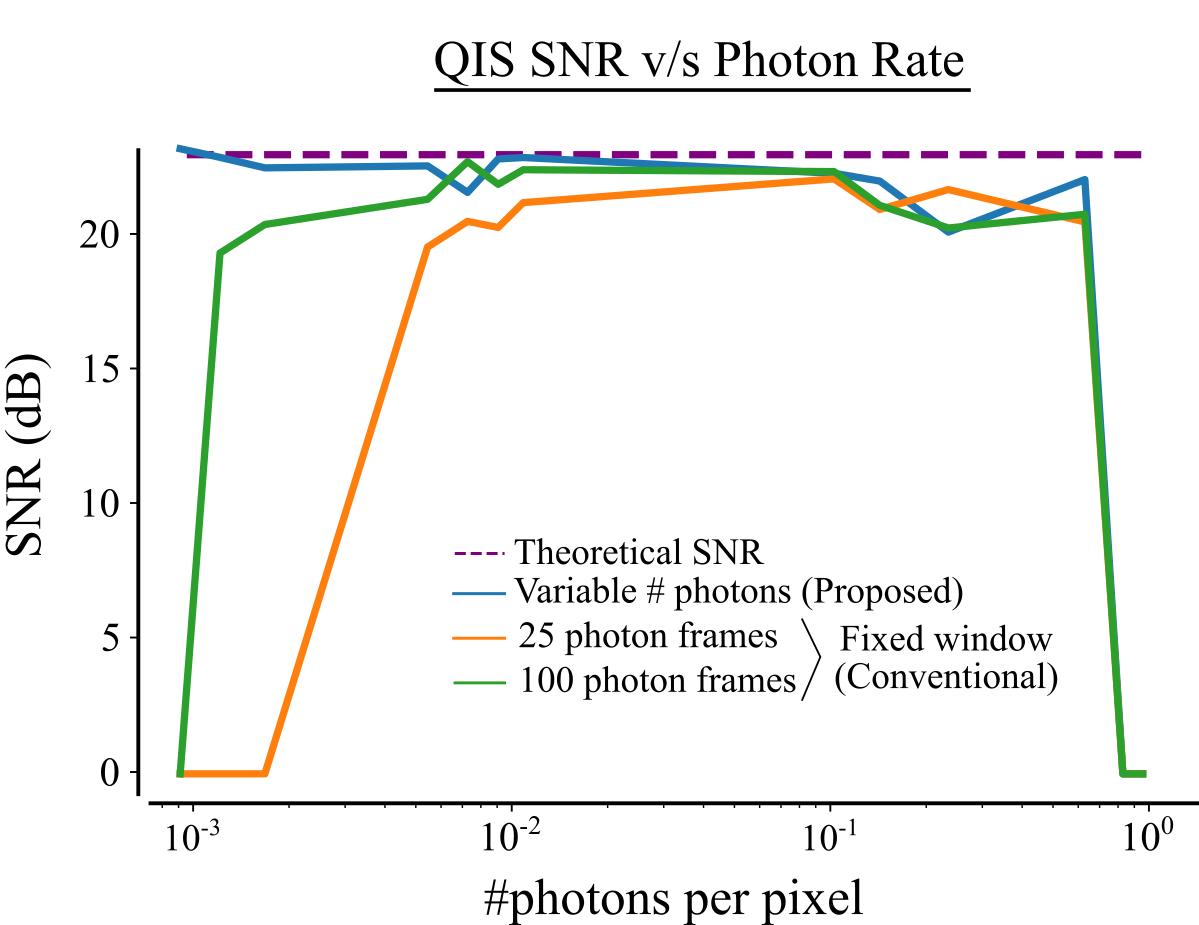}
  \caption{
   \textbf{QIS, SNR vs. Brightness.} We run the same experiment as in Fig. \ref{fig:PhotonBurstSize}, with the rotating orange at many light levels. We remove timing information from the raw data to create QIS binary data frames and run the same deblurring experiment. Again our adaptive changepoint method is helpful in low light scenarios. Note that the sharp drop on the right is due to saturation of a QIS sensor. 
    \label{fig:suppl_SNR_qis}}
\end{figure}

We run the same experiment as in Fig. \ref{fig:PhotonBurstSize}, where an orange is rotated at different brightness levels. We test our offline QIS changepoint detection method by removing the timing information and only considering a binary output. Our adaptive changepoint method helps at low light levels, see Fig. \ref{fig:suppl_SNR_qis}.  

\clearpage
\section{Additional Simulation Results}\label{suppl:additional_results}

This section contains some additional simulated results.  A second scene with
two toy cars is displayed, we use the same parameters as the toy car scene in
the main text for frame generation, changepoint detection and deblurring. For
this scene the dark car moves 90 pixels and has a contrast 3.3. The bright car
has a contrast of 1.2 and moves 30 pixels. Our results are shown in
supplementary Fig.~\ref{fig:suppl_sim_cars} and the clustered changepoints are
shown in supplementary Fig.~\ref{fig:suppl_clustered_changepoints}. Again, our
method is able to deblur both moving cars.

\begin{figure*}[!ht]
  \centering \includegraphics[width=\linewidth]{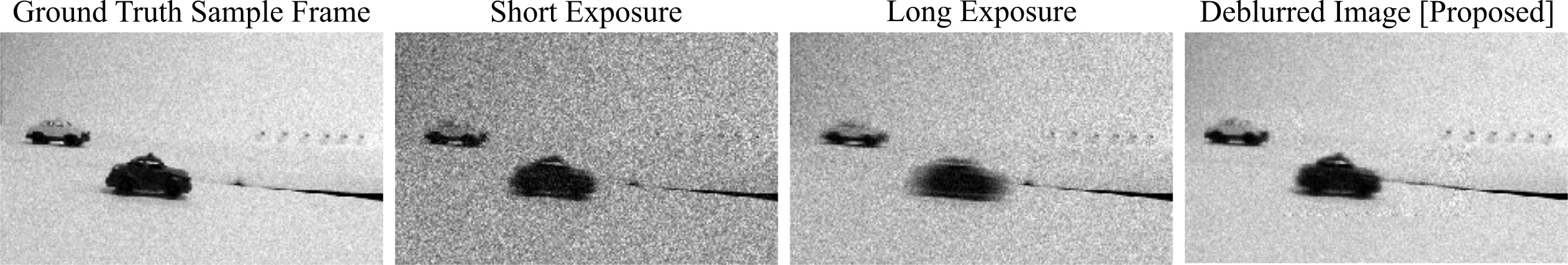}
  \caption{
    \textbf{Simulated Multiple Objects.} We simulate SPAD data from a 240 fps phone video of two rolling toy cars, a fast moving dark car and slow moving bright car. From left to right, the ground truth image shows the result of generating the same number of photon frames from the first frame of the video sequence. The short and long exposure images show the results of using only the first 75 and 250 photon frames, respectively. Notice that the short exposure preserves the dark car while the bright car is quite noisy, on the other hand, the long average blurs the dark car but preserves details of the bright one better. Finally, our deblurring algorithm is able to reconstruct both the dark and bright car.
    \label{fig:suppl_sim_cars}}
\end{figure*}

\begin{figure}[!ht]
  \centering \includegraphics[width=.6\columnwidth]{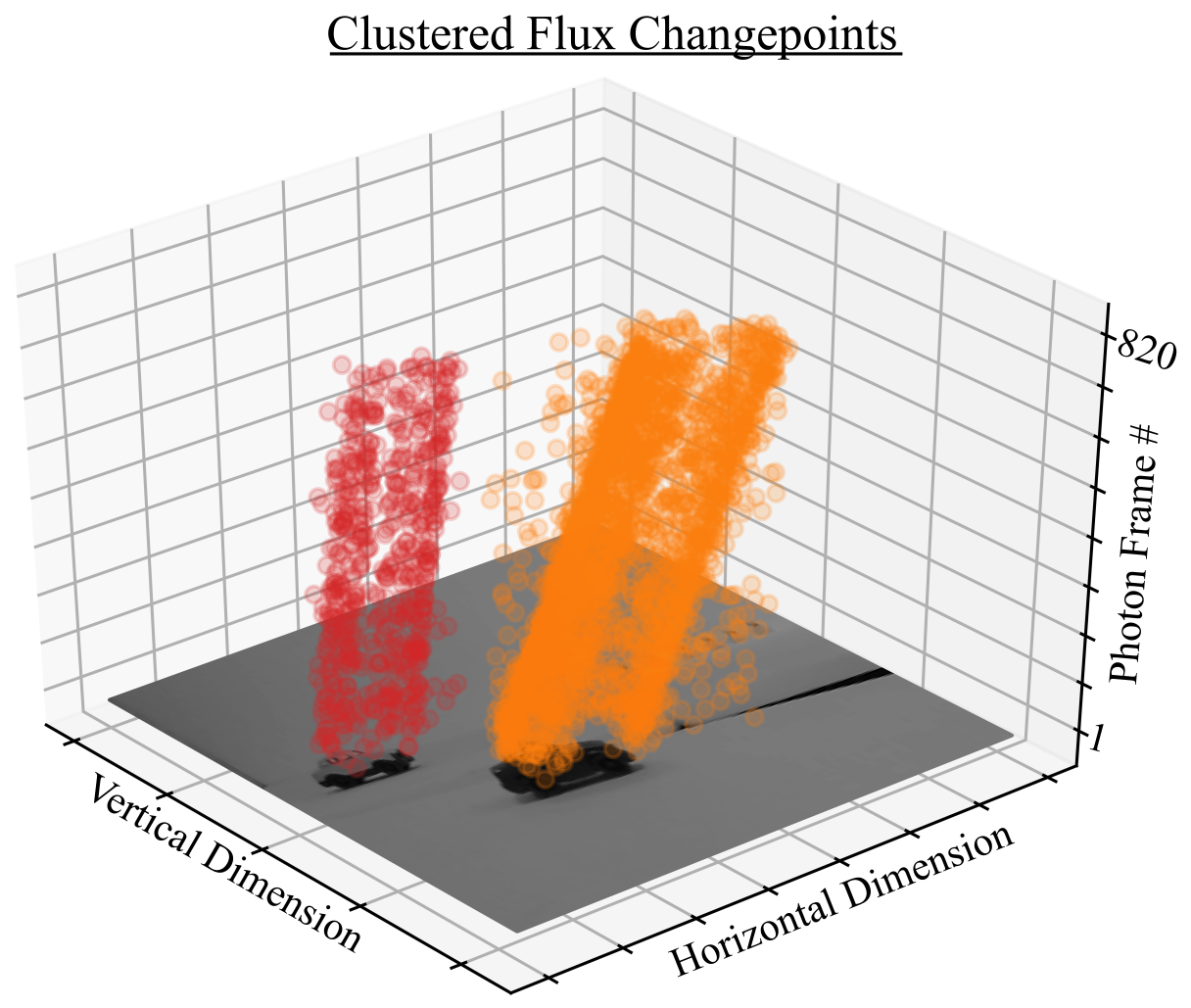}
  \caption{
    \textbf{Clustered Flux Changepoints} Displayed are the 2 flux changepoint clusters found for the scene in Supplementary Fig. \ref{fig:suppl_sim_cars}. We only display half of the flux changepoints in each cluster for visualization purposes.
    \label{fig:suppl_clustered_changepoints}}
\end{figure}

\newpage
We simulate a simple pixel art scene to demonstrate the advantage of deblurring on a changepoint video rather than burst frames. Notice in supplementary Fig. \ref{fig:suppl_pixel_cars} that the changepoint video frame is able to capture a wide range of motion speeds and contrasts that a single fixed frame cannot capture.

\begin{figure*}[!ht]
  \centering \includegraphics[width=.9\linewidth]{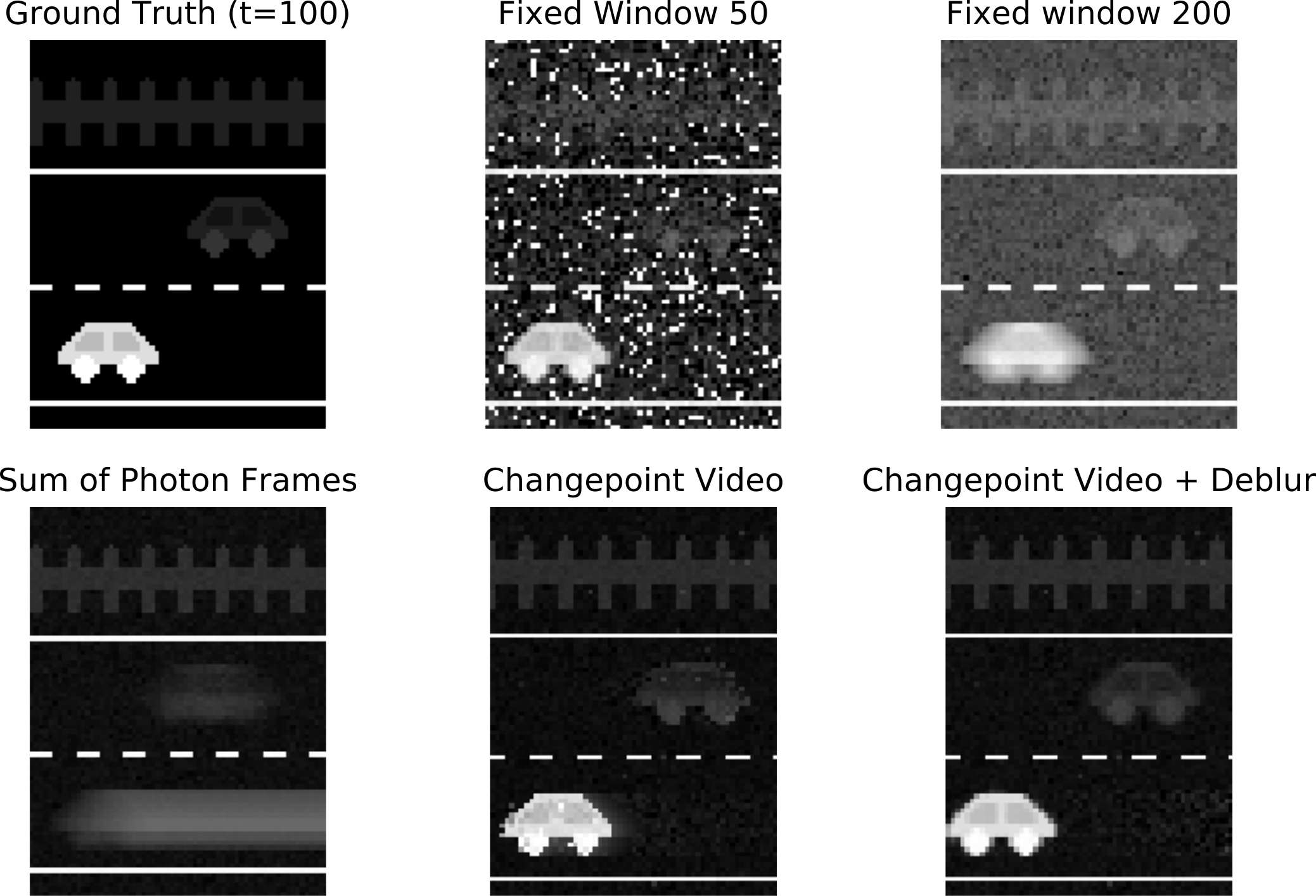}
  \caption{
   \textbf{Pixel Art Multiple Objects.} We simulate a scene where a bright car moves quickly to the right and a dim car moves slowly to the left. Notice that in the short averaging window, the dim car is lost in the noise while in the long averaging window the bright car is blurred. The sum of all photon frames maintains the background quite well. The changepoint video frame adapts to motion in each pixel and captures both the bright car, the dim car, and the background. Notice that the changepoint video frame loses some of the structure of the dim car due to noisy changepoints. We combine the adaptive changepoint video with a deblurring algorithm to deblur both cars.
    \label{fig:suppl_pixel_cars}}
\end{figure*}

\clearpage

\subsection{Global Motion Results}
\begin{figure*}[!htb]
\centering
\includegraphics[width=.8\linewidth]{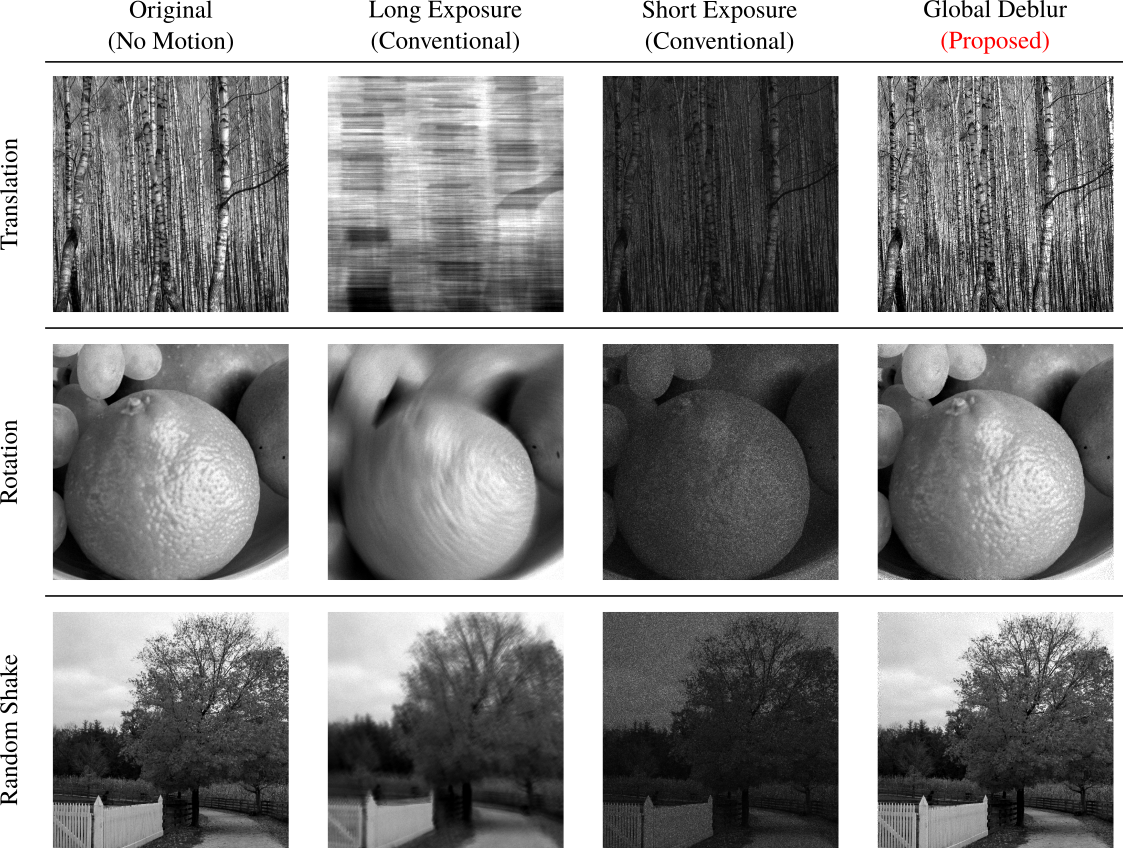}
\caption{\textbf{Simulated motion deblurring results for different types of global
motion.} From left the right the columns show, a
ground truth image that shows the result if the same number of photons are
sampled from the original image but with no motion. A long exposure where all
photon frames are summed, the result is a blurred
image. A short exposure image shows the combination of the first 20 photons of
the timestamp data, notice the edges are sharp but noise dominates. The result
of the global motion deblur algorithm is shown in the last column. (Original images from FreeImages.com)
\label{fig:suppl_resultSimulation}}
\end{figure*}

We start with a ground truth high resolution image, successively apply rigid transformations (using known rotations and translations), and generate photon frames using exponentially distributed arrival times. We reassign these high spatial resolution timestamps a lower resolution 2D array to simulate a low resolution SPAD pixel array. 

We model a photon frame readout from a SPAD array with \num{8000} bins and bin width of \SI{256}{\ps}. Images are scaled so that the true photon flux values
ranges between $10^4$ and $10^8$ photons per second. We then iteratively
transform the flux image according to known motion parameters, and downsample
spatially to a resolution of 425$\times$425 before generating photon timestamps.

For the horizontal translation blur, we moved the image 1 pixel to the right
blur, we rotate the image 0.1 degree for every 10 generated photons for a total
of 1000 photons. To emulate random camera shake, we create random motion
trajectories by drawing two i.i.d. discrete uniform random variables between
$-3$ and $3$ and use that as the number of pixels to translate along the
horizontal and vertical directions.  We generate 20 photons per translation for a total of 2000 photons.  We use the \textsc{BottomUp} algorithm
\citesuppl{rupturesPaper} with $\lambda=5$ for the changepoint detection step. In
practice we found that the results were not very sensitive to the choice of
$\lambda$ and values between $2$ and $12$ produced similar results. 

We generate photon events from an exponential distribution. We transform the
flux image, then down-sample to simulate objects with more detail than pixel
resolution. We then generate 10-20 photons from the down-sampled flux image.
Continuing this we get a 3-d tensor of photons representing global motion of the original image. 

\begin{figure}[!ht]
  \centering \includegraphics[width=0.5\linewidth]{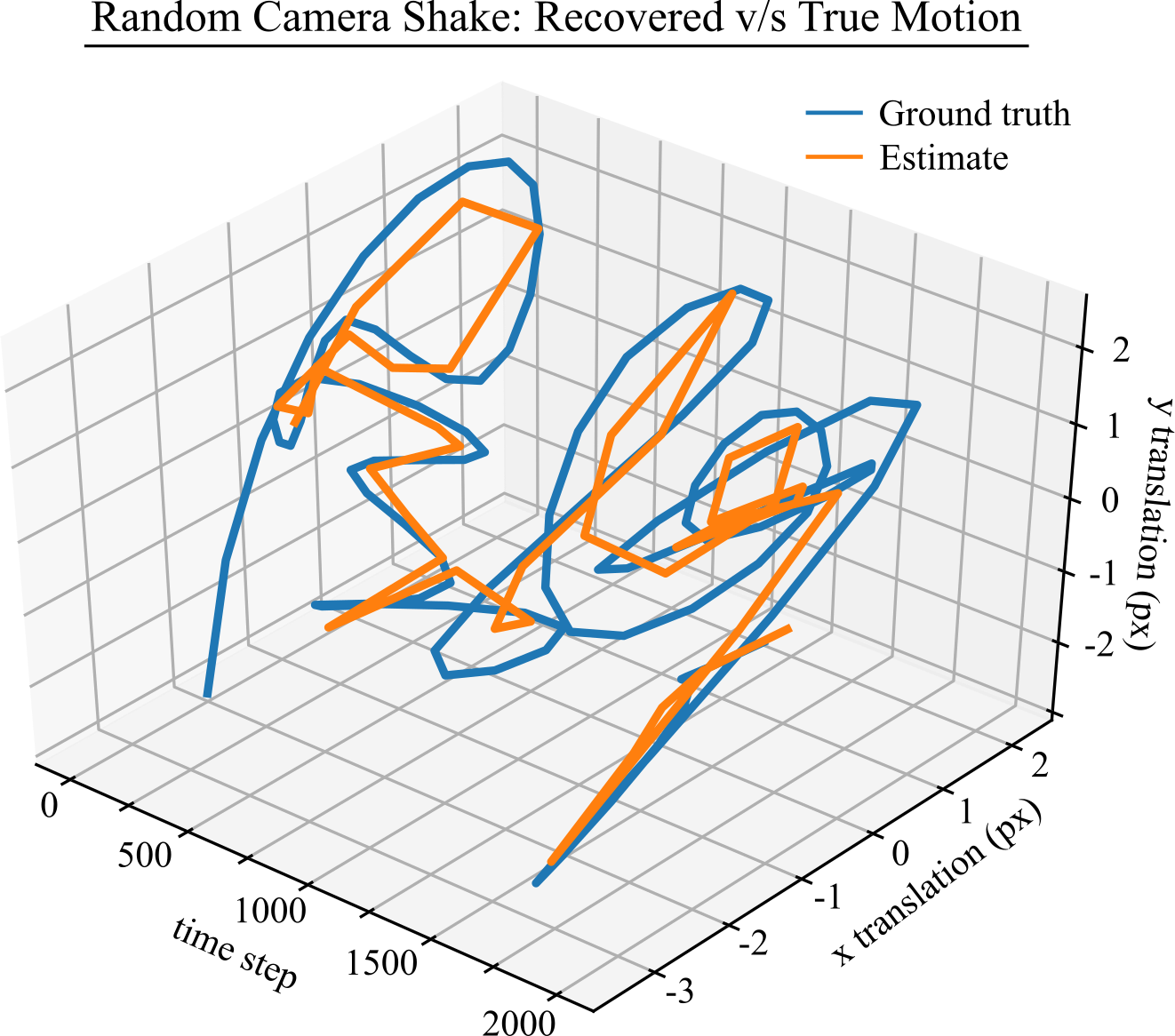}
  \caption{ \textbf{Comparison of estimated and true motion trajectories.}
    This plot shows the true and estimated motion trajectories for the random shake
    case in Fig.~\ref{fig:suppl_resultSimulation}. The recovered motion tracks the ground
    truth motion quite well. \label{fig:suppl_MotionTrajectoryComparison} }
\end{figure}

Supplementary Fig.~\ref{fig:suppl_resultSimulation} shows simulated deblurring results for three
different motion trajectories. The top row shows a case of horizontal translation: conventional long/short exposures must trade off motion blur and
shot noise.  Our deblurring method reproduces sharp details, such as vertical lines of the tree stems. The second row shows a case of rotation: note that different pixels of the scene now undergo different amount of motion per unit time. Our method reconstructs fine details of the texture of the orange peel. The bottom row shows random camera shake with unstructured motion. Our technique is able to correct for this global motion by approximating the overall motion trajectory as a sequence of small translations and rotations. Supplementary Fig.~\ref{fig:suppl_MotionTrajectoryComparison} shows the comparison between the true motion trajectory and the trajectory estimated as part of our deblurring algorithm.

\clearpage
\section{Comparison of BottomUp and PELT}\label{suppl:bottomup_v_pelt}
\begin{figure}[!ht]
  \centering \includegraphics[width=0.7\linewidth]{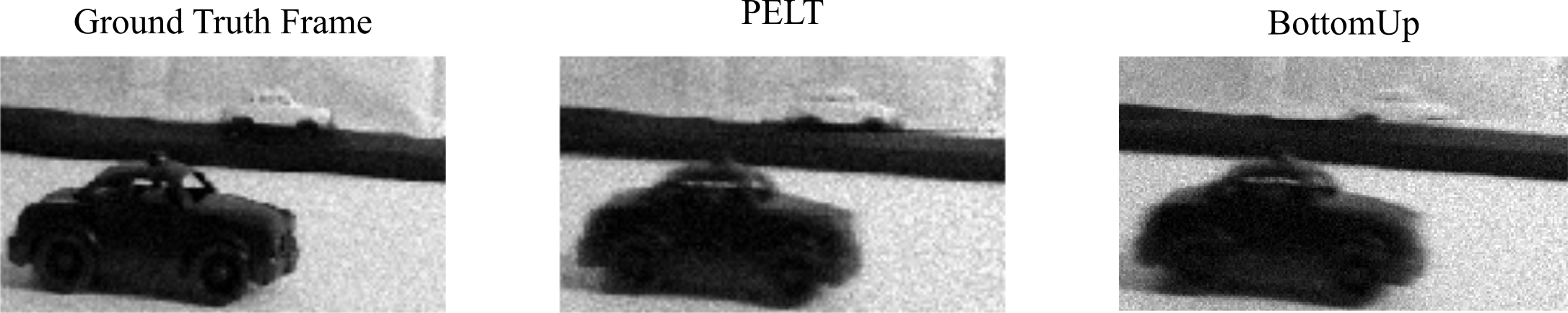}
  \caption{ \textbf{Toy Car Simulated Scene.}
    Here are the results using the toy car simulated scene. The parameters used are the same as in the main text. Note that the BottomUp algorithm is able to detect and deblur both cars; however, it seems to produce a slightly noisier and blurrier result. For this scene 210 by 300 with 690 photon frames per pixel, on our unoptimized system with 30 parallel cores, it takes the PELT algorithm two minutes to run while the BottomUp takes approximately one minute.  \label{fig:bottomup_vs_pelt_2cars} }
\end{figure}

\begin{figure}[!ht]
  \centering \includegraphics[width=.6\linewidth]{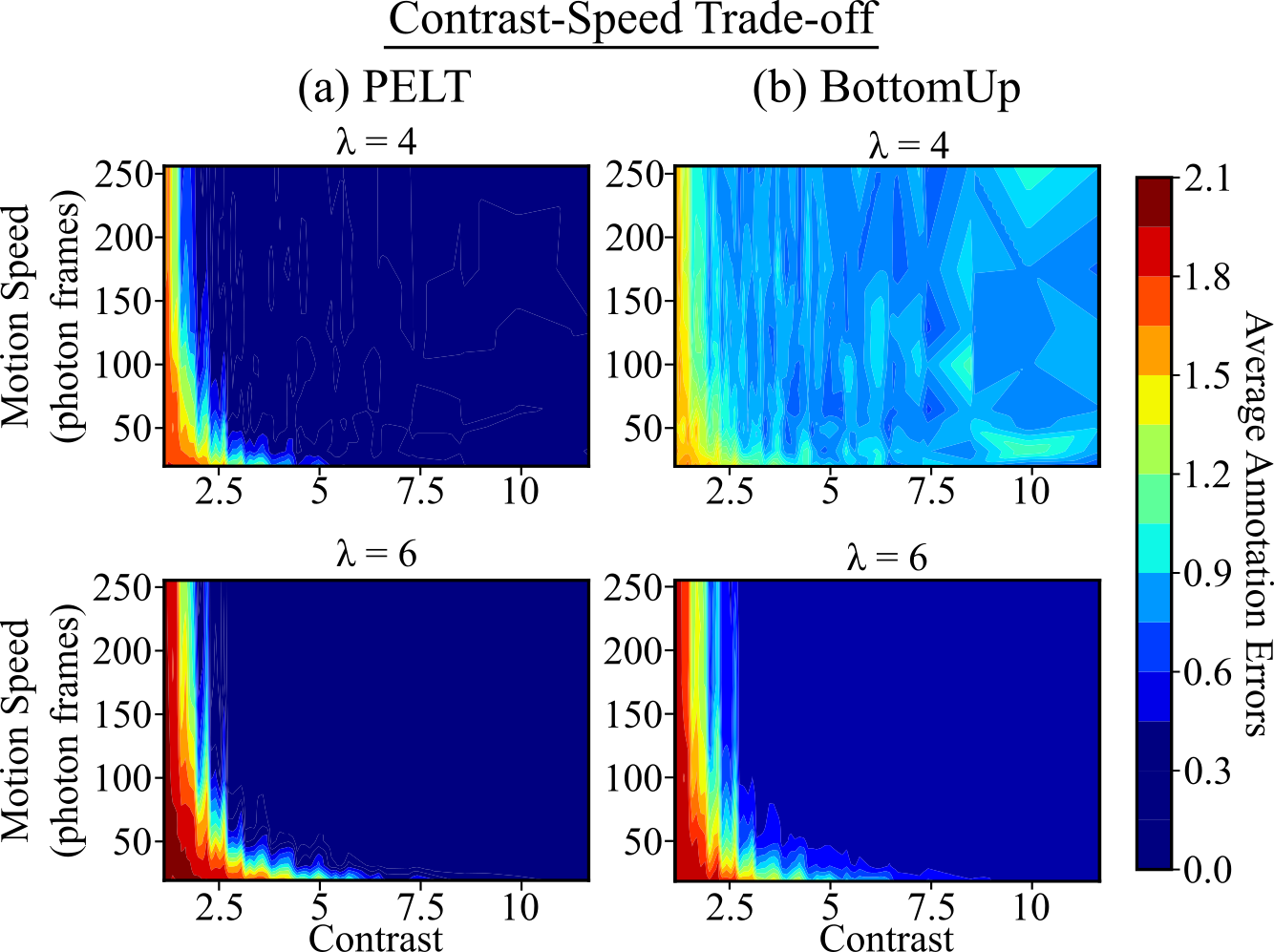}
  \caption{ \textbf{BottomUp vs. PELT - Contrast vs. Speed.}
  We repeat the contrast vs. speed simulation as done in the main text but with the BottomUp algorithm. Note that the BottomUp algorithm does comparably well for $\lambda=6$. For $\lambda=4$, the BottumUp algorithm is able to detect more difficult objects at the expense of false positives during easier scenarios. 
     \label{fig:bottum_up_vs_pelt_contrast_speed} }
\end{figure}

In this section we run some of the same simulation experiments from the main text but with the BottomUp algorithm instead of the PELT algorithm. In Suppl. Fig.~\ref{fig:bottomup_vs_pelt_2cars} we compare the ability of both algorithms on the toy car scene, BottomUp seems to produce slightly more noisy and blurry results. In Suppl. Fig.~\ref{fig:bottum_up_vs_pelt_contrast_speed} we re-run the contrast vs. speed simulations and find that BottomUp does comparably well with slightly more false positives.

\clearpage
\section{Experiment Setup}
For the experimental investigation we used a 32$\times$32 InGaAs SPAD array
from Princeton Lightwave  (PL GM-APD 32 x 32 Geiger-Mode Flash 3-D      LiDAR
Camera) and an RGB camera (VIS/BW point gray Grashopper 3, GS3-U3-23S6M-C)
capturing the same field-of-view. The SPAD camera samples photon events with a
depth of $10^{13}$ bins and \SI{250}{\ps/bin}. Further, during the experiments
we used a frame readout rate of \SI{50}{kHz}. The InGaAs sensor is sensitive in
near infrared (NIR) to shortwave infrared (SWIR) wavelengths ranging from
\SI{900}{\nm} to \SI{1.6}{\micro\meter}.

\begin{figure}[!h]
     \centering
     \centering \includegraphics[width=0.7\linewidth]{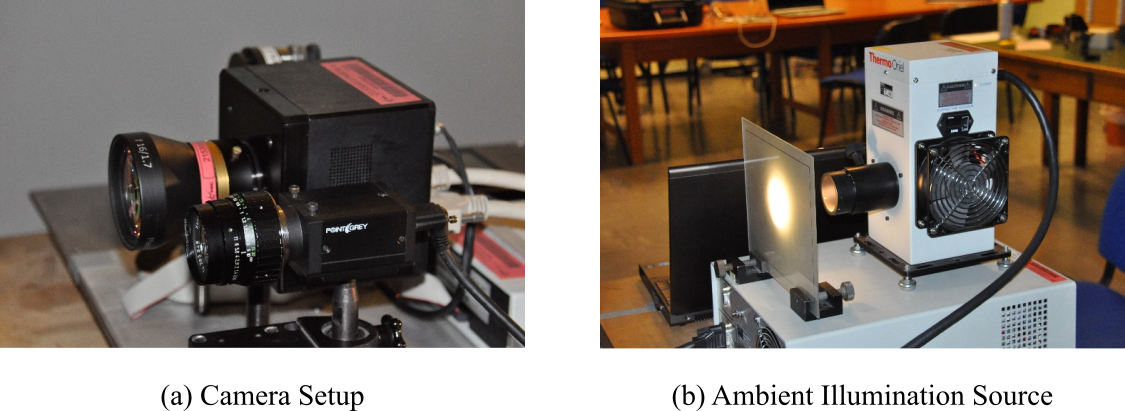}
     \caption{{\bf Hardware setup} (a) Our hardware setup consists of a
     Princeton Lightwave SPAD array (PL GM-APD 32$\times$32 Geiger-Mode Flash 3-D
     LiDAR Camera) and an RGB camera (VIS/BW point gray Grashopper 3,
     GS3-U3-23S6M-C) capturing the same field-of-view. (b) Ambient illumination
     is provided by a diffuse light source (broadband arc lamp ThermoOriel Model
     66881).
     \label{fig:expt_setup}}
 \end{figure}

During the measurements we investigated two different type of scene setups: a
``fan'' and a ``checkerboard'' scene, as depicted in
Suppl.~Fig.~\ref{fig:expt_setup}. In the first scene, the fan consists of three
blades mounted on a central cone and is enclosed by a circular frame with a
diameter of \SI{18}{\cm}. One blade was marked with a black piece of paper. The
fan scene was used to investigate rotational motion. The second scene consists
of an artificial head, a white plate and a colored checkerboard. Colors appear
at a different gray levels in the SWIR wavelength images.  This second scene
was used to investigate random motion due to a horizontally shaking camera.
 \begin{figure}[!h]
     \centering
     \centering \includegraphics[width=0.7\linewidth]{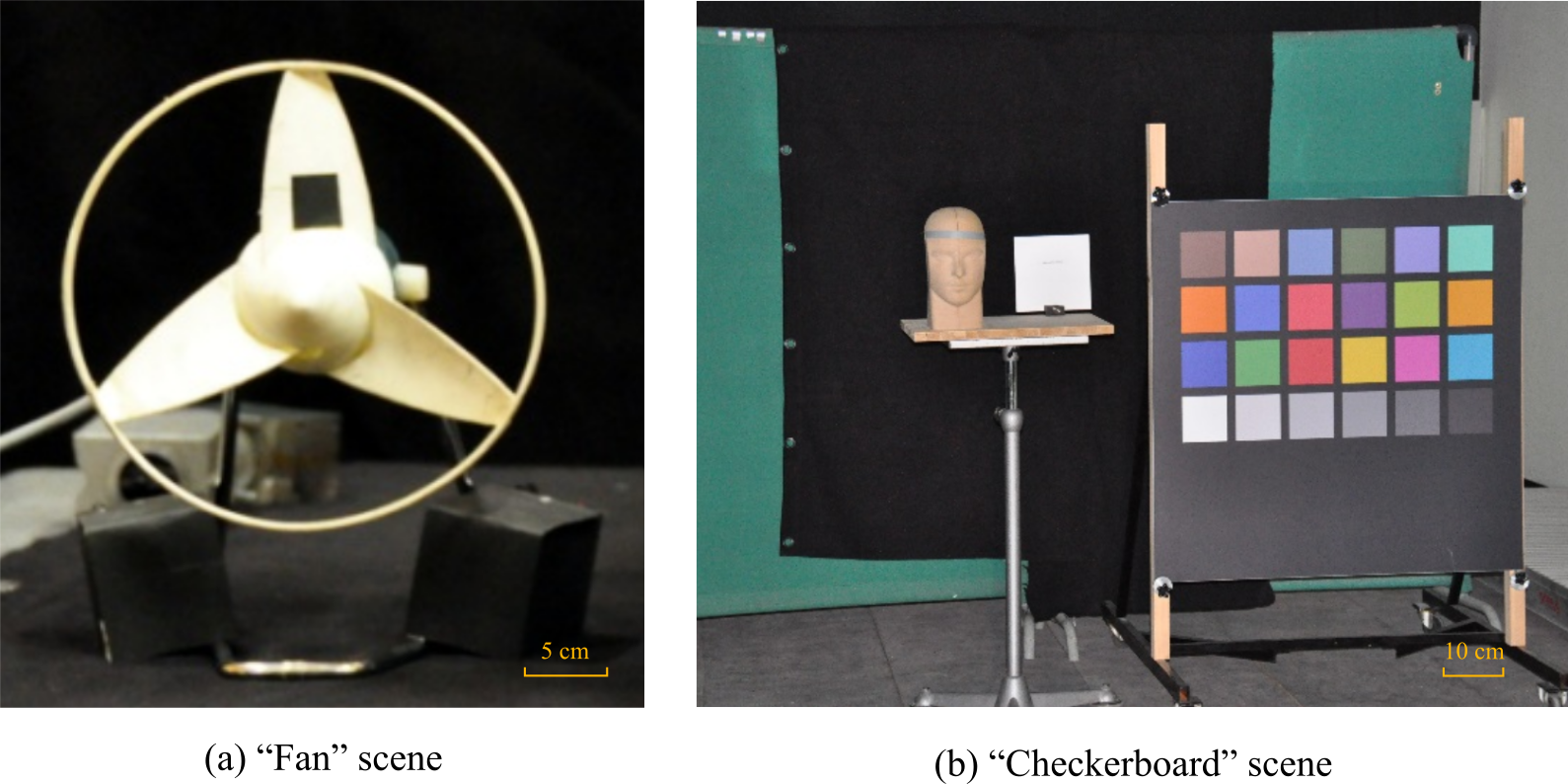}
     \caption{{\bf Experimental Scenes} (a) The ``fan'' scene consists of a small fan with
     a black square patch on one of the fan blades. (b) The ``checkerboard'' scene consists of
     a large color checkerboard and a mannequin head.
     \label{fig:expt_scenes}}
 \end{figure}

\section{Description of Video Results}
Please refer to included \texttt{.txt} and \texttt{.mp4} files for
supplementary video results.
%
%

\clearpage
{\small
\bibliographystylesuppl{ieee_fullname}
\bibliographysuppl{egbib}
}

\end{document}